# High-Mobility Indium Native Oxide Transistors via Liquid-Metal Printing in Air

*Shi-Rui Zhang*,[1], Sanjoy Kumar Nandi[1], Felipe Kremer[1], Shimul Kanti Nath[2], Wenzhong Ji[3], Thomas Ratcliff[1], Li Li[4], Nicholas J. Ekins-Daukes[2], Teng Lu[3], Yun Liu[3], and Robert Glen Elliman*,[1]*

[1]Department of Electronic Materials Engineering, Research School of Physics, The Australian National University, Canberra, ACT 2601, Australia

[2]School of Photovoltaic and Renewable Energy Engineering, University of New South Wales (UNSW Sydney), Kensington, NSW, 2052 Australia

[3]Research School of Chemistry, The Australian National University, Canberra, ACT 2601, Australia

[4]Australian National Fabrication Facility, Research School of Physics, The Australian National University, Canberra, ACT 2601, Australia

*Corresponding authors: Shi-Rui Zhang; Robert Glen Elliman

Email: shirui.zhang@anu.edu.au; rob.elliman@anu.edu.au

**ABSTRACT:** Oxide semiconductors have emerged as common channel materials in transistors and hold promise for next-generation electronics, yet achieving high mobility typically requires costly vacuum-based techniques. Here, ultrathin (5-nm) indium native oxide ($InO_x$) prepared by ambient-air liquid-metal printing (LMP) at low temperature (250 °C), is applied as semiconducting channel in field-effect transistor (FET). The resulting $InO_x$ is found to be polycrystalline with large lateral grains that extend vertically throughout the film thickness. $InO_x$ FETs in a transfer length method (TLM) configuration demonstrate a high conductivity mobility ($\mu_{CON}$) of 125 $cm^2\ V^{-1}\ s^{-1}$, with systematic analysis of contact resistance confirming potential for channel length scaling. Integration with atomic-layer-deposited (ALD) gate dielectrics further reveals excellent compatibility, for instance, $InO_x$ FET integrated with $HfO_2$ exhibits a high field-effect mobility ($\mu_{FE}$) of 107 $cm^2\ V^{-1}\ s^{-1}$, an on/off current ratio ($I_{ON}/I_{OFF}$) of $>10^7$, a subthreshold swing ($SS$) of 204 mV $dec^{-1}$, a gate leakage of $<10^{-6}$ A $cm^{-2}$, while maintaining stable performance over $10^4$ endurance cycles without degradation. Post-fabrication oxygen-plasma treatment is applied to achieve enhancement-mode operation and a depletion-load inverter is demonstrated, exhibiting a voltage gain of 69.8 V/V. These results demonstrate the great potential of LMP $InO_x$ as semiconducting channel in high-performance and power-efficient transistors for next-generation oxide electronics.

## INTRODUCTION

Since the invention of the thin-film transistor (TFT), oxide semiconductors have emerged as common channel materials due to their high carrier mobility, low leakage current and low process temperature. [1] Among them, amorphous indium gallium zinc oxide (a-IGZO) TFT stands out as a leading candidate and has been commercialized in active matrix for liquid-crystal displays (LCDs) and organic-light-emitting diode (OLED) display technologies. [2, 3] Recent advances in next-generation oxide electronics, particularly low-temperature (<400 °C), back-end-of-line (BEOL) compatible transistors, have renewed the interest in oxide semiconductors, [4-6] with indium oxide, a wide-bandgap oxide semiconductor, attracting great attention due to its high electron mobility of up to 160 $cm^2\ V^{-1}\ s^{-1}$. [7]

In transistor applications, indium oxide exists in either amorphous or polycrystalline phases. Owing to the s-orbital symmetry of indium and absence of grain boundary issues, amorphous indium oxide can retain high mobility. [8] Nevertheless, achieving an amorphous phase is challenging due to its strong crystallization tendency, even at low temperatures. To overcome this, atomic layer deposition (ALD) has been utilized to deposit amorphous indium oxide films with sub-nanometer thickness at a low thermal budget of 225 °C, where the extreme thickness scaling suppresses the crystallinity. Transistors based on these films exhibit high field-effect mobility ($\mu_{FE}$) exceeding 100 $cm^2\ V^{-1}\ s^{-1}$ and show great potential for three-dimensional BEOL integration. [9-12] On the other hand, polycrystalline indium oxide, often considered limited by grain boundary scattering, can also achieve high mobility when lattice ordering is improved and disorder-induced subgap states are minimized. For example, a high $\mu_{FE}$ reaching 139.2 $cm^2\ V^{-1}\ s^{-1}$ has been demonstrated in hydrogenated polycrystalline indium oxide ($In_2O_3$: H) FETs with enlarged grain size via a solid-phase crystallization (SPC) process at 300 °C.[13] Despite these advances, these approaches for depositing indium oxide remain dependent on costly vacuum-based techniques, motivating the search for low-cost and simple alternatives.

Liquid-metal printing (LMP) has recently emerged as a promising route for synthesizing metal native oxides in a vacuum-free, low-temperature, and cost-effective manner. [14-16] LMP oxides have already enabled a wide range of applications, including flexible electronics, [17] gas sensing, [18] photodetection, [19, 20] non-volatile memory, [21] logic circuit, [22-25] and neuromorphic computing. [26-28] Importantly, the ability to achieve controlled, large-area oxide nanosheets with lateral dimensions exceeding hundreds of millimeters, [29, 30] highlights the scalability of this method. In the context of field-effect transistors (FETs), LMP metal native oxides have been explored as both gate dielectrics and semiconducting channels. For examples, materials such as amorphous gallium oxide (a-$Ga_2O_3$) [24, 31, 32] and antimony oxide ($\alpha$-$Sb_2O_3$) [33] have been demonstrated as gate dielectrics and encapsulations for 2D transistors. Semiconducting channels based on variety of oxides have been reported, including tin oxides (SnO, $SnO_2$), [25, 34, 35] gallium oxide ($\beta$-$Ga_2O_3$), [23, 36, 37] tellurium dioxide ($\beta$-$TeO_2$), [38] $\alpha$-$Bi_2O_3$, [39] pristine indium oxide ($InO_x$, $In_2O_3$), [40, 41] indium tin oxide (ITO), [22] indium gallium oxide (IGO), [42] zinc-doped indium oxide (IZO), [43] antimony-doped indium oxide (IAO, AIO), [44, 45] and Indium gallium zinc oxide (IGZO). [46] Among these, p-type $\beta$-$TeO_2$ FETs exhibited record-high $\mu_{FE}$ of 232 $cm^{-2}$ $V^{-1}$ $s^{-1}$ at room temperature, while n-type indium oxide and zinc-doped indium oxide FETs exhibited $\mu_{FE}$ up to 96 $cm^{-2}$ $V^{-1}$ $s^{-1}$ and 87 $cm^{-2}$ $V^{-1}$ $s^{-1}$, respectively. The high mobility in p-type $\beta$-$TeO_2$ has been attributed to hybridization between Te 5s states and O 2p orbitals, leading to a dispersive valence band, whereas the high electron mobility in n-type oxide semiconductors typically originates from a dispersive conduction band contributed by spatially extended metal ns orbitals. [47]

Although LMP indium oxide FETs demonstrated promising high mobilities, key challenges still remain in device scaling. In particular, contact resistance can significantly impact device performance and lead to underestimated mobility values, [48] but this aspect has not been systematically studied in LMP indium oxide FETs. Moreover, indium oxide inherently possesses a high carrier density due to native oxygen vacancies, [49] making it difficult to effectively modulate the channel conductance through gate bias. This issue is further exacerbated in current reported LMP indium oxide FETs, which are often fabricated based on silicon dioxide ($SiO_2$) gate dielectric having thickness of several hundred nanometers. Due to the low capacitance density of thick $SiO_2$, a high gate voltage is required to perform a field-effect modulation. Low-voltage operation can be enabled by using high-capacitance gate dielectrics. [50, 51] While the use of thin high-dielectric-constant (high-$\kappa$) dielectrics, aluminium oxide ($Al_2O_3$) and hafnium oxide ($HfO_2$), particularly, enables better electrostatic control, [52] the compatibility of LMP indium oxide with these ALD high-$\kappa$ gate dielectrics have not yet been systematically investigated. Demonstrating such gate dielectric scaling is critical to reduce the operating voltage and promote the adoption of LMP technique in low-power transistors.

In this work, ultrathin indium native oxide ($InO_x$) nanosheet, prepared by ambient-air LMP technique at low temperature (250 °C), was applied as a semiconducting channel in FETs. Comprehensive material characterizations were conducted to investigate properties of the $InO_x$ nanosheet, including roughness, thickness, bandgap, and crystallinity. The fabricated $InO_x$ FETs were found to exhibit high conductivity mobility reaching 125 $cm^2$ $V^{-1}$ $s^{-1}$, where the impact of contact resistance was evaluated to assess the potential for channel length scaling. Furthermore, LMP $InO_x$ FETs integrated with ALD high-$\kappa$ gate dielectrics, such as $Al_2O_3$ and $HfO_2$, were systematically investigated to assess the potential for equivalent oxide thickness (EOT) scaling. Device performance was evaluated in terms of mobility, subthreshold swing ($SS$), gate leakage, and interface trap density. Notably, the $InO_x$ FET with $HfO_2$ gate dielectric achieves high mobility of 107.2 $cm^2\ V^{-1}\ s^{-1}$ and an $SS$ value as low as 204.3 mV $dec^{-1}$, while maintaining stable performance without degradation over $10^4$ endurance cycles. Post-fabrication treatment with oxygen plasma was applied to achieve enhancement-mode operation and a depletion-load inverter was demonstrated.

## RESULTS AND DISCUSSION

FETs with $InO_x$ semiconducting channels were fabricated using LMP technique in ambient air, which involved a pressure-assisted process to exfoliate $InO_x$ nanosheets from molten indium onto a substrate, as depicted in Figure 1a. The substrate was heated to 250 °C, and a molten indium droplet was applied, during which it formed a self-limiting $InO_x$ surface layer. A second preheated substrate was then pressed uniaxially onto the molten indium, spreading it between the substrates and facilitating the transfer of the $InO_x$ nanosheets. After printing and cleaning, the $InO_x$ nanosheet remained adhered to the substrate due to strong van der Waals interaction. The FET channel was then defined using lithography and inductively coupled plasma (ICP) etching. The devices were completed by adding source and drain electrodes using standard metal deposition and lift-off processes (see the Methods section for details).

Indium oxide crystallizes at relatively low temperatures, forming a stable body-centered cubic (BCC) bixbyite structure (space group *I*a3) with a lattice parameter of approximately 1.0117 nm. [53] However, as the thickness of indium oxide decreases to the single- or sub-nanometer range, degree of crystallinity decreases and the amorphous phase becomes more stable. [54] Atomic force microscopy (AFM) of the LMP $InO_x$ nanosheet in this study shows a thickness of 5 nm and a root-mean-square (RMS) roughness of 0.45 nm (Figure 1b). The level of roughness is comparable to that of thermally grown $SiO_2$, which exhibits an RMS roughness of 0.30 nm, and helps reduce surface scattering during carrier transport. The crystallinity of the nanosheet was first investigated by Grazing Incidence X-ray diffraction (GIXRD) (Figure 1c). This reveals dominant diffraction peaks at 30.8°, 35.6°, 51.1°, 60.9°, corresponding to the (222), (400), (440), (622) planes of BCC indium oxide, respectively (See COD card No.2310009). The optical bandgap of the nanosheet was determined from ultraviolet–visible transmittance measurements (Figure 1d) by extrapolating the linear region of a Tauc plot to the energy axis (Inset in Figure 1d). This reveals a direct bandgap ($E_g$) of 3.65 eV, consistent with values reported for $In_2O_3$ and its role as a wide-bandgap oxide semiconductor.

The thickness and crystallinity of the $InO_x$ nanosheet were confirmed by cross-sectional transmission electron microscopy (TEM) (Figure 1e). Nano-beam diffraction (NBD) of the $InO_x$ region shows that a single-crystalline structure extends through the full thickness of the $InO_x$ nanosheet, and that it has a body-centered cubic bixbyite structure. Diffraction pattern obtained from the Pt protection layer and the $SiO_2$ substrate reveals the nanocrystalline nature of Pt and the amorphous structure of $SiO_2$, neither of which contributes to the diffraction signal of $InO_x$ (Figure S1). Corresponding energy-dispersive X-ray spectroscopy (EDX) was used to further confirm the elemental composition of the same region (Figure S2).

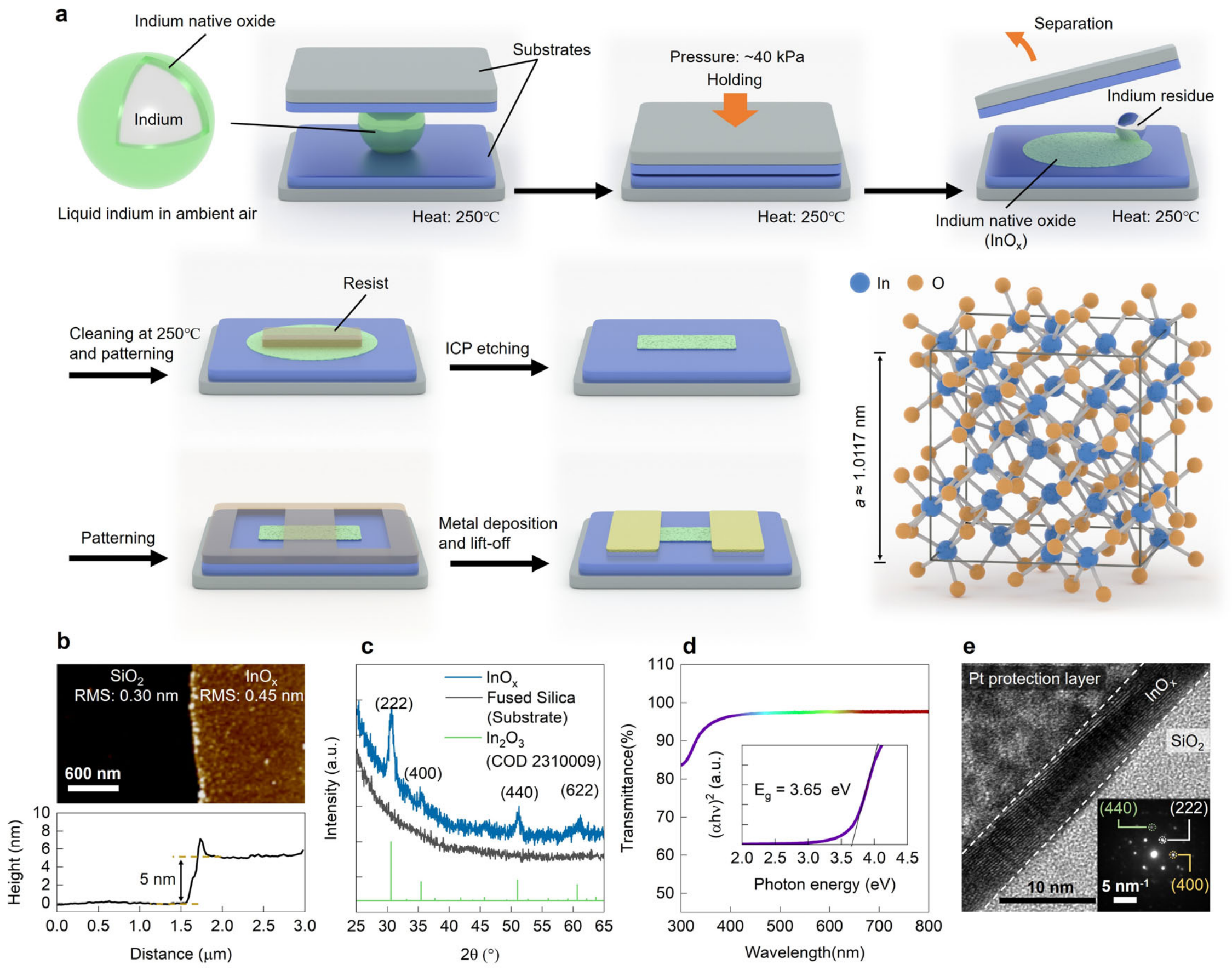

**Figure 1.** $InO_x$ as semiconducting channel in FETs and materials characterization. (a) Schematic illustration of device fabrication based on $InO_x$ through pressure-assisted liquid-metal printing and rendered crystal structure of cubic indium oxide. Crystal structure was rendered based on the data from the Crystallography Open Database.[55] (b) AFM height image of $InO_x$ nanosheet on $SiO_2$ and corresponding step-height profile. (c) GIXRD scans of $InO_x$ nanosheet and fused silica substrate. (d) UV-Vis transmission spectrum of $InO_x$ nanosheet. Inset: Tauc plot and extracted optical bandgap. (e) Cross-sectional TEM image of $InO_x$ nanosheet. Inset: Indexed diffraction pattern of $InO_x$ region.

Cross-sectional TEM and SAED verify the through-film (out-of-plane) crystallinity of the $InO_x$ nanosheet but plan-view (in-plane) analysis is also required to fully characterize the film microstructure. Because the crystallinity of LMP oxides are affected by applied pressure and the compliance of the substrate, [23] it is important to perform such analysis on actual device structures. This was achieved using the approach shown in Figure 2a. A series of backside polishing and etching steps was undertaken on the $InO_x/SiO_2/Si$ structure to gradually expose the $SiO_2$ layer. The amorphous $SiO_2$ was then etched to a thickness less than 100 nm, which was sufficient for TEM analysis (see Methods for details).

Figure 2b shows a plan-view TEM image of the polycrystalline $InO_x$ nanosheet, clearly revealing grain boundaries and grains. The enlarged view shown in Figure 2c displays well-defined lattice fringes, corresponding to the (222) plane of cubic indium oxide. SAED was also used to confirm the polycrystalline nature of $InO_x$ nanosheet on amorphous $SiO_2$, where 5 typical diffraction rings are prominent, corresponding to (211), (222), (400), (440), (622) planes, as shown in Figure 2d. The lateral grain size in the $InO_x$ nanosheet was determined by AFM in-phase imaging (Figure 2e). The size distribution for 100 grains is shown in Figure 2f, and reveals an average lateral grain size of 23.5 ± 6.8 nm. These results confirm that the $InO_x$ nanosheet possesses a polycrystalline structure with large lateral grains that extend vertically throughout the film thickness, which is distinct from the previously reported LMP indium oxide films that exhibit a layered polycrystalline structure. [40] The through-film crystallinity may reduce charge trapping associated with internal boundaries or defects and improve vertical electron transport.

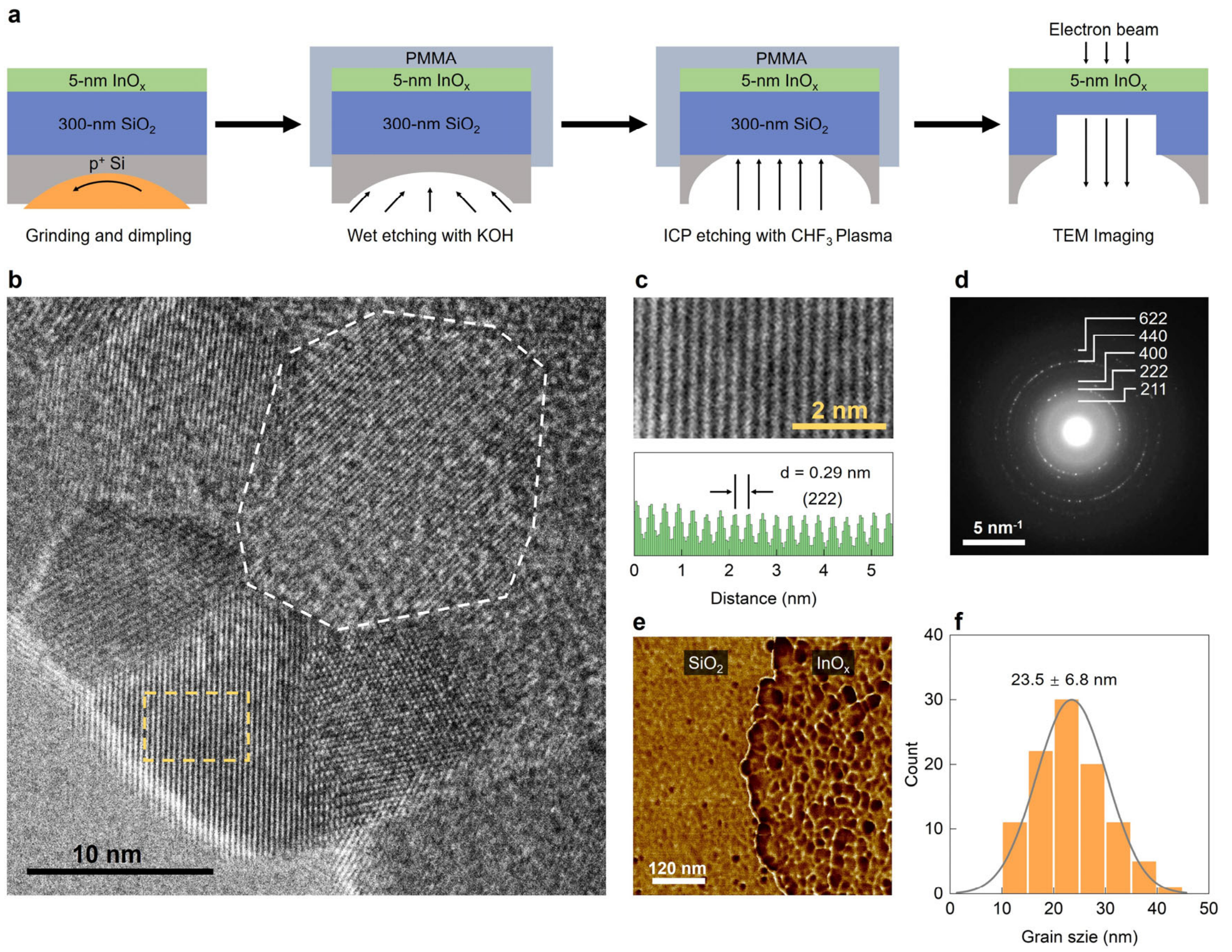

**Figure 2.** Lateral crystallinity of $InO_x$ nanosheet. (a) Schematic illustration of preparing sample for plan-view TEM imaging. (b) TEM image of $InO_x$ nanosheet on $SiO_2$ with a typical grain highlighted by a dashed polygon. (c) Magnified view of the lattice pattern within a single grain marked by a dashed square in (b) and a direct lattice-spacing measurement from the fringes. (d) SAED pattern of $InO_x$ nanosheet on $SiO_2$. (e) AFM in-phase image of $InO_x$ nanosheet on 300-nm $SiO_2$. (f) Grain size distribution of $InO_x$ nanosheet based on the measurement in (e).

The electrical performance of $InO_x$ FETs was evaluated using the transfer length method (TLM), as shown in Figure 3a. The gate dielectric is a 300-nm thermal $SiO_2$ layer with an oxide capacitance $C_{OX}$ = 11.4 nF cm$^{-2}$, as determined from a capacitance-voltage (CV) measurement of a fabricated p$^+$ Si/$SiO_2$/Ni/Au capacitor (Figure S3). Structural parameters such as channel length ($L_{CH}$), channel width ($W_{CH}$) and contact length ($L_C$) were confirmed using scanning electron microscopy (SEM) (Figure 3b).

Drain current versus gate-source voltage ($I_D$-$V_{GS}$) measurements were performed on FETs with different $L_{CH}$ in a single TLM structure, using drain-source voltage ($V_{DS}$) of 0.1 V with $V_{GS}$ sweeping from -150 V to 150 V. The large value of $V_{GS}$ is necessary to fully modulate the conductivity of $InO_x$ on 300-nm $SiO_2$, due to the relatively low capacitance density of the latter. $V_{DS}$ is sufficiently small to ensure linear-regime operation and that the carriers in the semiconducting channel are uniform from source to drain.

Figure 3c shows transfer ($I_D$-$V_{GS}$) curves of n-type FETs with $L_{CH}$ of 0.6 μm, 1μm, 2 μm, 3 μm and 6 μm. Threshold voltage ($V_{TH}$) can be extracted from the extrapolation of the linear portion of the transfer curves, and the carrier density near the source can be estimated as $n_S = C_{OX}V_{OV}/q$, where $V_{OV}$ is the overdrive voltage ($V_{OV} = V_{GS} - V_{TH}$, for n-channel FETs) and q is the elementary charge.[56] Figure 3d shows the width-normalized total resistance ($R_{TOT}$) of FETs as a function of $L_{CH}$, as extracted from the linear region of the transfer curves at $V_{OV}$ ranging from 80 V to 150 V. While the transfer length ($L_T$) can be extracted from the x-intercept of these plots, the contact resistance ($R_C$) can be extracted from the y-intercepts according to $R_{TOT} = 2R_C + R_{SH}L_{CH}$, where $R_{SH}$ is the sheet resistance.

Figure 3e shows the extracted values of $R_C$ and $L_T$ as a function of $n_S$. The reduction in $R_C$ with increasing $n_S$ is attributed to the effect of contact gating, where the Schottky barrier (SB) is modulated by the back-gate voltage. High back-gate voltages reduce the width of the SB and facilitate the injection of carriers, thereby reducing $R_C$. $R_C$ is estimated to be ~2.0 kΩ μm at $n_S \approx 1.07 \times 10^{13}$ cm$^{-2}$. $R_C$ is slightly higher than advanced transistors using ALD amorphous indium oxide, which may due to a higher measured surface roughness of LMP polycrystalline $InO_x$ and extra residues or interfacial inhomogeneity introduced from additional LMP process steps. The contact resistance could therefore likely be reduced by improving the cleanliness of the metal/oxide interface. For example, a mild plasma could be used to remove residues, though its impact on the electrical properties of $InO_x$ would need to be carefully evaluated; It could be further reduced by increasing the carrier density in the contact region. On the other hand, $L_T$ remains constant at different $n_S$ and is estimated to be ~429.5 nm, confirming sufficient carrier injection from the metal contact with $L_C$ of 1 μm in TLM device. What should be noted is that contact doping can effectively convert part of the nominal channel near the Ni/$InO_x$ interface into an $n^+$ region with a lower sheet resistance, thereby shortening the effective channel length. Under such conditions, the TLM analysis here based only on long-channel FETs ($L_{CH} \geq 0.6$ μm) can underestimate the transfer length, because the total resistance may no longer scale linearly with the nominal channel length assuming a single sheet resistance. Accurate assessment of length of the doped $n^+$ region and its sheet resistance requires including sufficiently short-channel FETs ($L_{CH} \leq 80$ nm), [57] enabling a more reliable extraction of transfer length.

$R_{SH}$ is extracted from the slope of the $R_{TOT}$ versus $L_{CH}$ plot. As shown in Figure 3f, $R_{SH}$ decreases as $n_S$ increases, indicating an efficient field-effect modulation from the back gate. The conductivity mobility ($\mu_{CON}$), calculated based on $\mu_{CON}=1/qn_SR_{SH}$, reflects the intrinsic electrical properties of the $InO_x$ nanosheet under back-gate control without the impact of contact resistance. The calculated $\mu_{CON}$ reaches a high value of ~125.4 cm$^2$ V$^{-1}$ s$^{-1}$ with only a subtle dependence on back-gate voltage. Since the uncertainty in the $\mu_{CON}$ is determined by the fitting uncertainty of the sheet resistance, the standard error of the extracted sheet resistance is <1%, corresponding to <1% uncertainty in this calculated conductivity mobility value. The field-effect mobility ($\mu_{FE}$) is extracted from the highest value of transconductance and calculated according to $\mu_{FE} = L_{CH}g_m/(W_{CH}V_{DS}C_{OX})$ at $V_{DS} = 0.1$ V, where $g_m$ is the transconductance defined by $g_m = dI_D/dV_{GS}$. Low standard errors (<6%) of the extracted values indicate that the measured data closely follow the expected linear scaling with small residuals. This suggests good consistency among the FETs in the TLM structure and supports the robustness of the TLM extraction. The extracted $\mu_{FE}$ values of FETs with $L_{CH}$ of 6 μm, 3μm, 2 μm, 1 μm, 0.6 μm are 111.2, 98.5, 88.6, 71.4, and 51.2 cm$^2$ V$^{-1}$ s$^{-1}$, respectively (Figure S4). Although the presence of contact resistance leads to a noticeable reduction in $\mu_{FE}$ when reducing channel length, $\mu_{FE}$ remains above 50 cm$^2$ V$^{-1}$ s$^{-1}$ even at channel length of 600 nm. The combination of high mobility and outstanding performance in small-scale devices highlights the LMP $InO_x$ as a promising channel material for high-performance FETs and their further device scaling.

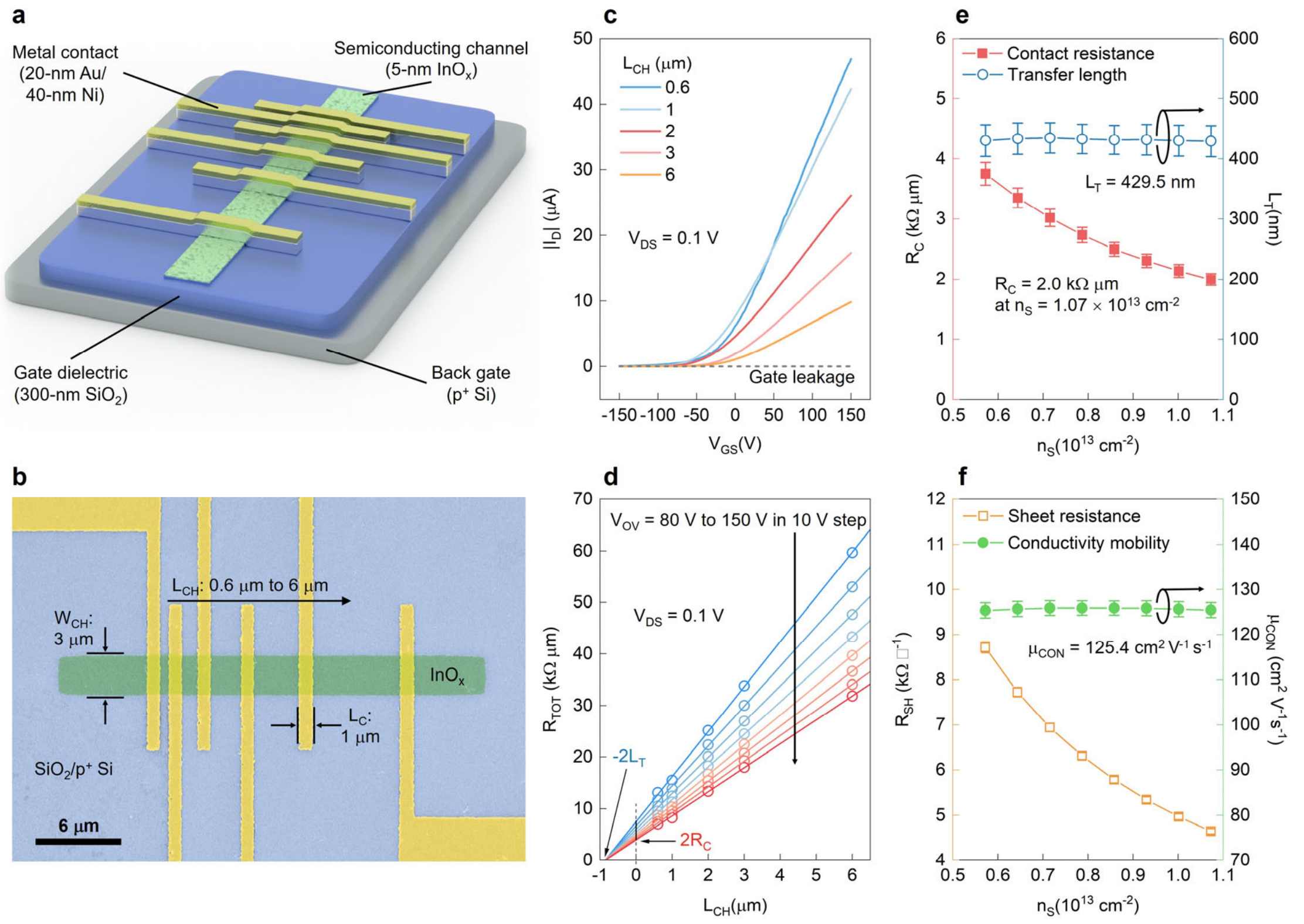

**Figure 3.** Back-gate TLM configuration and electrical characteristics. (a) 3D schematic illustration of back-gate TLM structure. (b) False-color SEM image of the representative fabricated TLM device. (c) Transfer ($I_D$-$V_{GS}$) curves in linear scale of $InO_x$ FETs with $L_{CH}$ ranging from 0.6 μm to 6 μm. (d) $R_{TOT}$ versus $L_{CH}$ and linear fittings under different $V_{OV}$ from 80 V to 150 V for extracting $2L_T$ and $2R_C$. (e) $R_C$ (left axis) and $L_T$ (right axis) versus $n_S$. (f) $R_{SH}$ (left axis) and $\mu_{CON}$ (right axis) versus $n_S$. In (e) and (f), error bars reflect the standard errors from the linear regression of $R_{TOT}$ versus $L_{CH}$.

Although $InO_x$ FETs with 300-nm $SiO_2$ gate dielectric have exhibited high mobility, they require high gate voltages to switch. To address this limitation, FETs were also fabricated with high-$\kappa$ gate dielectrics, where EOT scaling enables enhanced field-effect modulation. High-$\kappa$ $Al_2O_3$ and $HfO_2$ grown by plasma-enhanced atomic layer deposition (PEALD) were used. The electrical uniformity of LMP $InO_x$ layers on different gate dielectrics was confirmed by scanning capacitance microscopy (SCM) (Figure S5). The SCM dC/dV mapping visualizes the distribution of carriers induced by back-gate bias. Samples with high-$\kappa$ gate dielectrics show a higher contrast between the $InO_x$ channel region and the underlying dielectric region, indicating enhanced field-effect modulation.

Figure 4a shows a schematic of the device structure, together with an SEM image of a fabricated device defining the parameters: $L_C$, $L_{CH}$, $W_{CH}$, which in this case have dimensions of 6 μm, 6 μm, and 3 μm, respectively. A sufficiently long channel minimizes the influence of contact resistance on the extracted FET mobility, while an extended contact length ensures efficient carrier injection into the semiconducting channel. Figure 4b shows the typical transfer curves of representative $InO_x$ FETs using 300-nm $SiO_2$ as gate dielectrics. The subthreshold swing ($SS$) is extracted from the subthreshold region of the transfer curves according to $SS = dV_{GS}/d\log I_D$. The $InO_x$ FET with a 300-nm $SiO_2$ gate dielectric exhibits a minimum $SS$ of 7.4 V $dec^{-1}$ at $V_{DS}$ = 5 V and a high operating $V_{GS}$ of 150 V. In contrast, $InO_x$ FETs with 30-nm $Al_2O_3$ and 30-nm $HfO_2$ dielectrics show reduced operating $V_{GS}$ of 5 V and 3 V, hereafter will be called $InO_x/Al_2O_3$ FET and $InO_x/HfO_2$ FET, respectively.

Figure 4c and Figure 4g show typical transfer curves and field-effect mobilities of $InO_x$ FETs based on 30-nm $Al_2O_3$ and 30-nm $HfO_2$ gate dielectrics with capacitance densities of 210.9 nF $cm^{-2}$ and 429.4 nF $cm^{-2}$, respectively (Figure S3). Both FETs exhibit high $\mu_{FE}$ values of 107.5 $cm^2\ V^{-1}\ s^{-1}$ and 107.2 $cm^2\ V^{-1}\ s^{-1}$, respectively, along with on/off current ratios ($I_{ON}/I_{OFF}$) of more than $10^7$ under large $V_{DS}$. Figure 4d and Figure 4h display the output characteristics ($I_D$-$V_{DS}$) of the FETs, where ohmic behavior is observed at small $V_{DS}$ and current saturation is observed as $V_{DS}$ approaches $V_{GS}$-$V_{TH}$ due to the pinch-off effect. While these field-effect mobilities are derived from the transconductance, another important parameter, the effective mobility ($\mu_{eff}$), can be calculated based on the drain conductance ($g_d$) at small $V_{DS}$, according to $\mu_{eff} = L_{CH}g_d/(W_{CH}V_{OV}C_{OX})$. [58] $InO_x/Al_2O_3$ FET exhibits a high $\mu_{eff}$ of 104.0 $cm^2\ V^{-1}\ s^{-1}$ at $V_{GS}$ = 5 V, while that of the $InO_x/HfO_2$ FET is calculated to be 115.2 $cm^2\ V^{-1}\ s^{-1}$ at $V_{GS}$ = 3 V. High $\mu_{FE}$ values exceeding 160 $cm^2\ V^{-1}\ s^{-1}$ are also extracted from FETs based on unpatterned $InO_x$ channels (Figure S6). However, these values might be overestimated as a result of fringe effects, [59, 60] i.e., the current extends beyond the nominal channel width leading to an overestimation of the current density and an overestimated mobility. Due to a weak field-effect modulation outside the apparent channel area, higher off-state currents are also observed in the subthreshold regime, reducing $I_{ON}/I_{OFF}$ by at least on order of magnitude compared to FETs with patterned $InO_x$ channels.

Figures 4e and 4i compare the $SS$ of $InO_x/Al_2O_3$ and $InO_x/HfO_2$ FETs as a function of $I_D$, in the range of $10^{-11}$ A to $10^{-7}$ A. The $InO_x/Al_2O_3$ FET exhibits a minimum $SS$ of 313.1 mV $dec^{-1}$ at $V_{DS}$ = 5 V, while the $InO_x$ /$HfO_2$ FET has a lower minimum $SS$ of 204.3 mV $dec^{-1}$ at $V_{DS}$ = 3 V, which is attributed to the higher capacitance density of $HfO_2$. To investigate the quality of semiconductor/dielectric interface in these $InO_x$ FETs, the interface trap density ($D_{IT}$) is extracted using the equation $SS = \ln(10)(1 + C_S/C_{OX} + qD_{IT}/C_{OX})k_BT/q$, where $k_B$ is the Boltzmann constant, $T$ is the temperature, q is the elementary charge, and $C_S$ is the semiconductor capacitance (equal to zero for fully depleted FETs) [61]. $D_{IT}$ values of $5.56 \times 10^{12}\ eV^{-1}\ cm^{-2}$ and $6.47 \times 10^{12}\ eV^{-1}\ cm^{-2}$ are obtained for $InO_x/Al_2O_3$ FET and $InO_x/HfO_2$ FET, respectively. $SS$ could be further reduced here by EOT scaling using thinner gate dielectrics here, whereas approaching the thermal limit would require additional interface engineering and gate-dielectric engineering. [62] Finally, the gate leakage of $InO_x/Al_2O_3$ and $InO_x/HfO_2$ FETs are shown in Figure 4f and Figure 4j, respectively. Both FETs have gate leakage far below the low-power limit (leakage current density < $1.5\times10^{-2}$ A $cm^{-2}$), [63] which can be attributed to the excellent insulating properties of $Al_2O_3$ and $HfO_2$, and to the small contact area of channel/dielectric junction, which limits uni-directional gate current. [64] $I_D$-$V_{GS}$ measurements were also performed on 20 $InO_x/HfO_2$ FETs at $V_{DS}$ = 0.1 V (Figure S7). The extracted $\mu_{FE}$ shows a maximum of 110.4 $cm^2$ $V^{-1}$ $s^{-1}$, with an average of 88.0 ± 21.4 $cm^2$ $V^{-1}$ $s^{-1}$. The *SS* exhibits a minimum of 166.1 mV $dec^{-1}$, with an average of 231.1 ± 41.9 mV $dec^{-1}$. All the FETs have gate leakage current of lower than $10^{-5}$ A $cm^{-2}$.

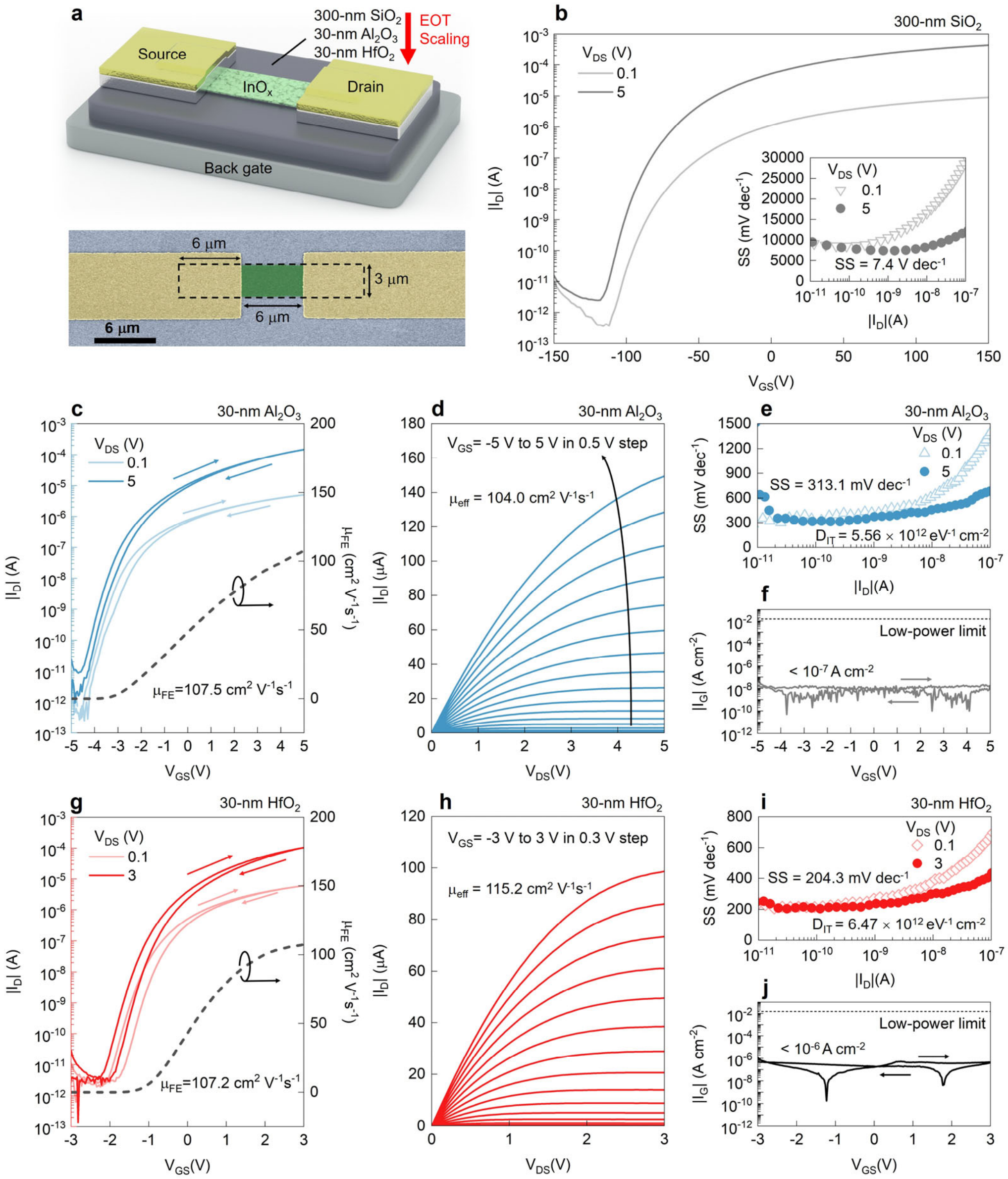

**Figure 4.** Back-gate FET device structure and electrical characteristics. (a) 3D schematic illustration of back-gate FET and a false-color SEM image of a representative fabricated FET. (b) Transfer ($I_D$-$V_{GS}$) curves in logarithmic scale of FET with 300-nm $SiO_2$ as gate dielectric. Inset: $SS$ versus $I_D$. (c-f) Transfer ($I_D$-$V_{GS}$) curves in logarithmic scale (left axis) and $\mu_{FE}$ versus $V_{GS}$ (right axis), Output ($I_D$-$V_{DS}$) curves, $SS$ versus $I_D$, Gate leakage ($I_G$-$V_{GS}$) of $InO_x/Al_2O_3$ FET. (g-j) Transfer ($I_D$-$V_{GS}$) curves in logarithmic scale (left axis) and $\mu_{FE}$ versus $V_{GS}$ (right axis), Output ($I_D$-$V_{DS}$) curves, $SS$ versus $I_D$, Gate leakage ($I_G$-$V_{GS}$) of $InO_x/HfO_2$ FET.

The performance of the representative LMP $InO_x/HfO_2$ FET in this work is benchmarked against other indium oxide and doped-indium oxide transistors fabricated by various techniques, such as sputter, ALD, pulsed laser deposition (PLD), and solution process (Sol) as shown in Figure 5a. Significantly, the LMP $InO_x/HfO_2$ FET, which is fabricated at a relatively low process temperature, demonstrates a device mobility comparable to transistors fabricated by ALD or sputter deposition. It also stands out as one of the highest-mobility FETs realized via vacuum-free processes.

$InO_x/HfO_2$ FET in this work also has the lowest gate dielectric EOT, less than 10 nm, while simultaneously maintaining one of the highest mobilities among reported n-type LMP transistors (Figure 5b). The mobility values reported in previous LMP indium oxide studies might have been underestimated because the devices typically employed large channel area ($W_{CH} \geqslant 40$ µm and $L_{CH} \geqslant 20$ µm), where squeeze-printed or pressure-assisted-printed nanosheets with lateral dimensions of tens of micrometers can potentially exhibit thickness non-uniformity or partial discontinuity (Figure S8). A representative FET device with partially interrupted channel shows an extracted $\mu_{FE}$ of 63.2 $cm^2\ V^{-1}\ s^{-1}$ when calculated using the nominal channel geometry (Figure S9). Further studies on printing process are required to improve the printing uniformity and ensure the structural integrity of large-area nanosheets. In this work, narrow channels were intentionally used to minimize geometric ambiguity arising from local thickness variations or partial discontinuities and to ensure reliable mobility extraction. Importantly, recent advances in scalable liquid-metal printing, such as dewetting-induced printing with a print head and continuous liquid-metal printing (CLMP) using a roller, [29, 30, 42] provide more controlled transfer conditions and rely on standard solvent cleaning to remove residual metal, which may improve thickness uniformity and reduce discontinuities over large areas.

When benchmarking power consumption, it is worth noting that EOT alone does not reflect the operating voltages of transistors due to differences in semiconducting channel and interface quality across studies. Instead, *SS* offers a more direct indication of the voltage required to switch a transistor. It is therefore a more relevant parameter for evaluating switching efficiency and power consumption, although it is not reported in all LMP transistor studies. For a fair comparison, the device mobility is also benchmarked versus *SS* values for those LMP transistor studies where such values were reported (Figure 5c). The $InO_x/HfO_2$ FET in this work remains among the top-performing n-type FETs, offering both high mobility and one of the lowest reported *SS* values.

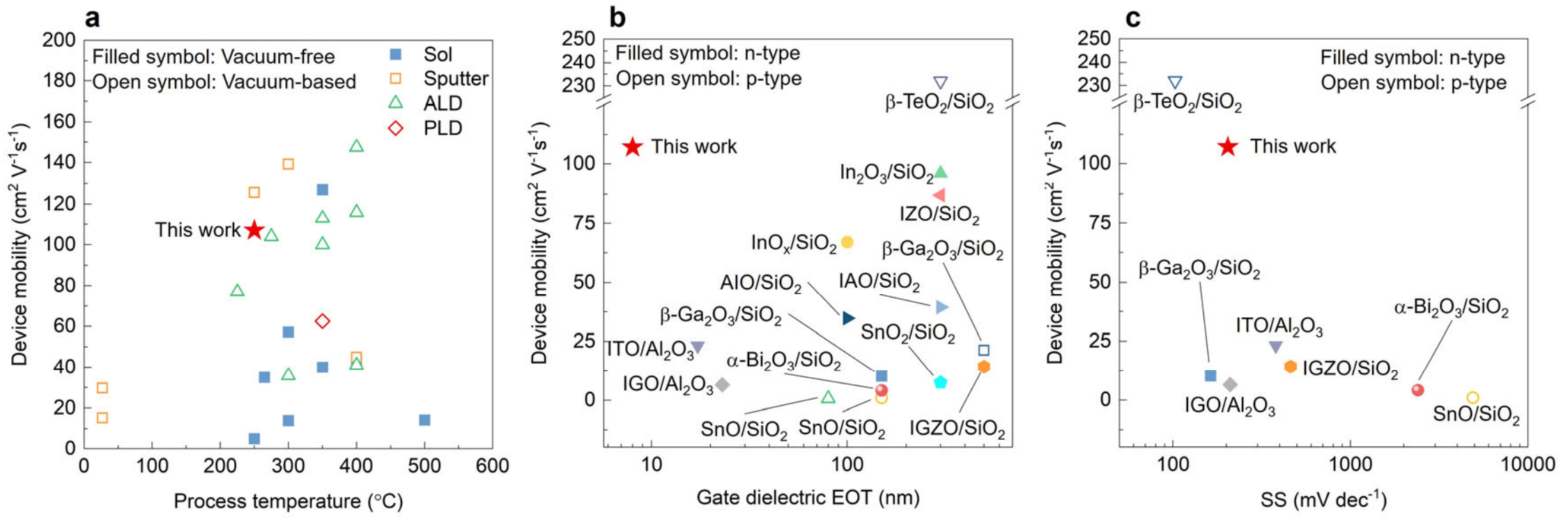

**Figure 5.** Benchmarking of $InO_x/HfO_2$ FET against other reported transistors. (a) Comparison of device mobility versus process temperature among indium oxide or doped-indium oxide transistors using different techniques. (b) Comparison of device mobility versus gate dielectric EOT among LMP transistors. (c) Comparison of device mobility versus *SS* among LMP transistors. The data used for comparison in this figure are listed in Table S1 and Table S2

To investigate the reliability of $InO_x/HfO_2$ FET, bias stressing tests and endurance test were carried out under ambient condition. The back-gate bias for positive bias stressing (PBS) and negative bias stressing (NBS) were +3 V and -3 V, respectively. Figure 6a and 6b shows changes in transfer characteristics under PBS and NBS for 5000 s. The FET exhibits a positive $V_{TH}$ shift of around +0.5 V under PBS, and recovers to its initial state within 1000 s after the bias stress is removed, as shown in Figure 6c. The positive $V_{TH}$ shift may originate from interface charge trapping. During PBS, electrons are gradually captured at the trap sites at dielectric interface, where they form an electrostatic screen against field-effect modulation and cause a positive shift of $V_{TH}$. Once the positive bias is removed, these electrons are gradually released, leading to a recovery of $V_{TH}$. The FET exhibits a negative $V_{TH}$ shift of around -0.8 V under NBS, and a similar recovery process is observed once the bias stress is removed, as shown in Figure 6d. The negative $V_{TH}$ shift is likely caused by water ($H_2O$) molecules adsorption on the semiconducting channel. During NBS, $H_2O$ molecules adsorbed on the channel surface trap holes and become positively charged, subsequently attracting electrons, which then accumulate in the channel, leading to a negative shift of $V_{TH}$.[65] After the bias stress is removed, $H_2O$ molecules are gradually desorbed, resulting a recovery of $V_{TH}$. However, further controlled-atmosphere measurements and encapsulation would be required to unambiguously identify this as the dominant mechanism. In practical applications, encapsulation would be necessary to suppress NBS-induced $V_{TH}$ shift under ambient conditions.

Figure 6e shows $10^4$ cycles of transfer sweeps, exhibiting stable switching performance of $InO_x/HfO_2$ FET under consecutive cycling. $I_{ON}$ and $I_{OFF}$ as well as $\mu_{FE}$ are extracted from each cycle (Figure 6f). Both $I_{ON}$ and $\mu_{FE}$ have a cycle-to-cycle variability of < 2%, while $I_{OFF}$ remains at a low level (~$10^{-13}$ A), with fluctuation that can be attributed to environment noise and resolution limit of measurement system.

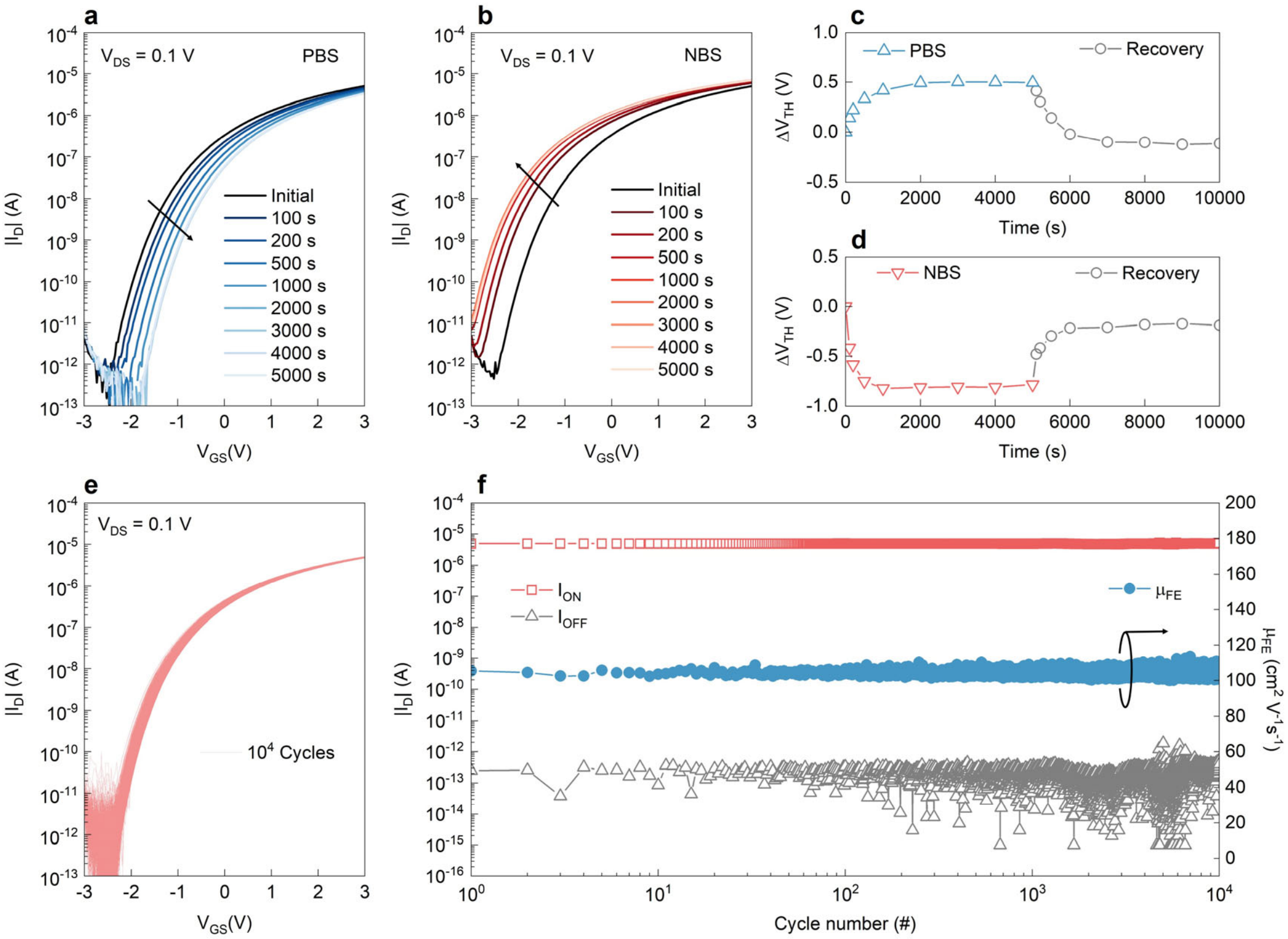

**Figure 6.** Reliability of $InO_x/HfO_2$ FET. Transfer curves under (a) PBS and (b) NBS for 5000 s. $V_{TH}$ shift during (c) PBS and recovery, (d) NBS and recovery. $V_{GS}$ values for PBS and NBS were +3 V and -3V, respectively. (e) Consecutive transfer curves for $10^4$ cycles. (f) Cycle-to-cycle variation of $I_{ON}$, $I_{OFF}$ and $\mu_{FE}$.

The as-fabricated $InO_x/HfO_2$ FETs operate in depletion mode due to high electron density in channel layer. Several methods have proved effective at shifting $V_{TH}$ positively by suppressing oxygen vacancies and thereby reducing electron density in indium oxide, such as oxygen annealing, [66] mild $CF_4$ plasma doping, [11] and oxygen plasma treatment. [67] To demonstrate the transition from depletion-mode (D-mode) to enhancement-mode (E-mode) operation, post-fabrication treatment with oxygen plasma was applied to as-fabricated FETs. The oxygen-plasma treatment caused negligible change in the surface roughness of $InO_x$ (Figure S10) and did not degrade the cycle-to-cycle stability (Figure S11). A slight reduction in field-effect mobility was observed, consistent with previous studies on plasma-treated indium oxide transistors, and may be related to a reduced effect carrier concentration after plasma treatment.[11, 67]

Figure 7a shows the transfer curves at $V_{DS}$ of 0.1 V of a representative $InO_x/HfO_2$ FET before and after oxygen-plasma treatment. The $V_{TH}$ shifts positively for more than 2 V and enhancement-mode operation is realized. Application as a depletion-load inverter (logical NOT gate) was demonstrated, where a D-mode FET serves as the load and an E-mode FET works as a drive. Figure 7b and figure 7c show voltage transfer characteristics (VTC) of the inverter and corresponding voltage gains, respectively, with supply voltage ($V_{DD}$) varied from 1 to 5 V. When the input voltage ($V_{IN}$) is low and close to zero ("logic 0"), output voltage ($V_{OUT}$) approaches $V_{DD}$ ("logic 1"), whereas when $V_{IN}$ is high and close to $V_{DD}$ ("logic 1"), $V_{OUT}$ falls to near zero ("logic 0"). Consistent with the logic operation, the corresponding voltage gain increases with $V_{DD}$ and reaches 69.8 V/V at $V_{DD}$ = 5 V.

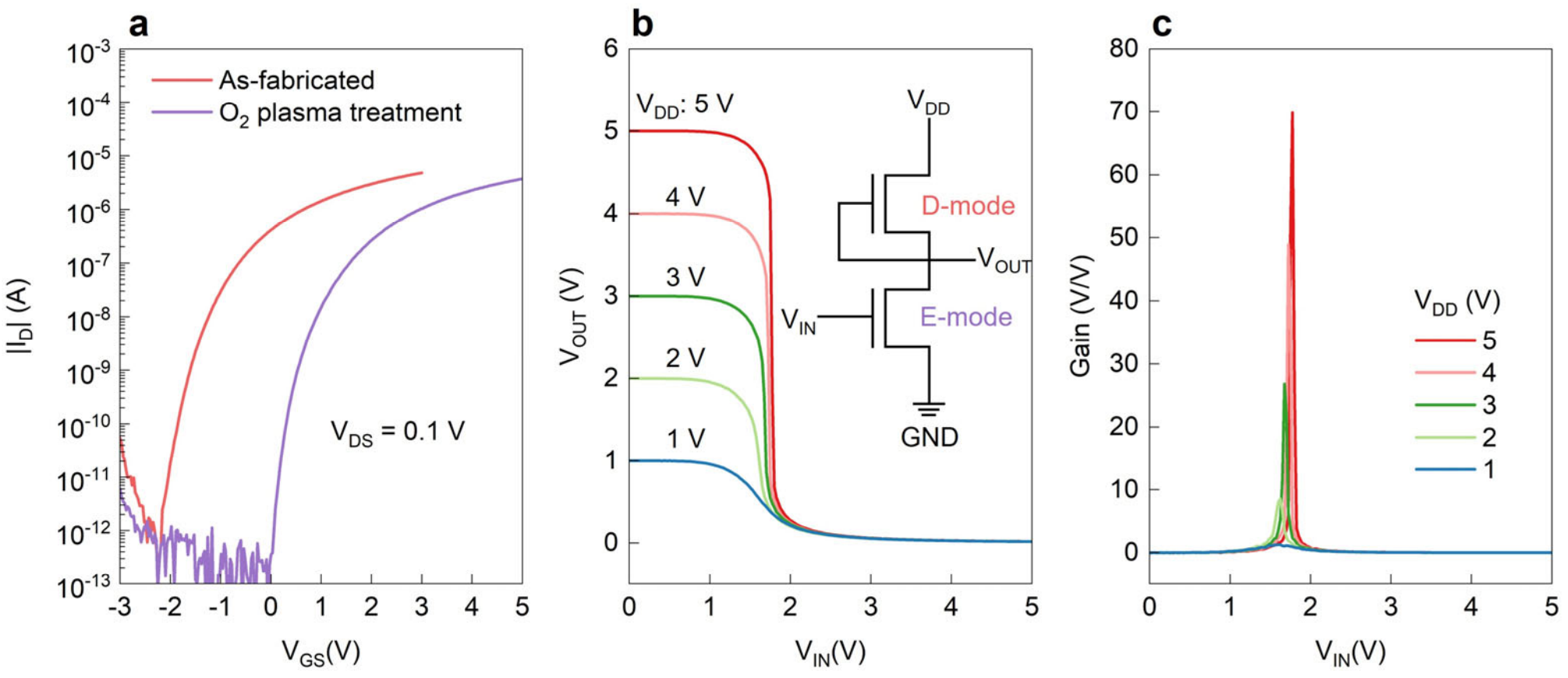

**Figure 7.** Post-fabrication treatment and depletion-load inverter. (a) Transfer curves of the as-fabricated depletion-mode FET and the $O_2$ plasma-treated enhancement-mode FET. (b) Voltage transfer characteristics (VTC) of the depletion-load inverter at varied $V_{DD}$ and equivalent circuit diagram. (c) Corresponding voltage gains of the depletion-load inverter at different $V_{DD}$.

## CONCLUSIONS

In this work, $InO_x$ nanosheet prepared by ambient-air LMP technique at low temperature of 250 °C was applied as a semiconducting channel in FETs. The $InO_x$ nanosheet exhibits a uniform thickness of 5 nm and an RMS surface roughness of 0.45 nm. The $InO_x$ is found to be polycrystalline with a body-centered cubic bixbyite structure, with the grains extending through the film thickness and having a mean lateral size of ~23.5 nm.

Back-gate $InO_x$ FETs in a TLM structure demonstrate $\mu_{CON}$ values as high as 125 $cm^2\ V^{-1}\ s^{-1}$, where the impact of $R_C$ is evaluated. The extracted $\mu_{FE}$ is shown to decrease with shrinking $L_{CH}$, due to the increasing impact of $R_C$, but remained above 50 $cm^2\ V^{-1}\ s^{-1}$ even at $L_{CH}$ of 600 nm, indicating significant potential in channel length scaling. Furthermore, LMP $InO_x$ FETs integrated with ALD high-$\kappa$ gate dielectrics, $Al_2O_3$ and $HfO_2$, was systematically studied to assess EOT scaling. The resulting FETs achieve $I_{ON}/I_{OFF}$ of $>10^7$ and mobility exceeding 100 $cm^2\ V^{-1}\ s^{-1}$, with *SS* values as low as 313.1 and 204.3 mV $dec^{-1}$ for $InO_x/Al_2O_3$ and $InO_x/HfO_2$ FETs, respectively. These devices also show low $D_{IT}$ values and gate leakage that meets the low-power limit requirement. In addition, $InO_x/HfO_2$ FET also exhibit excellent reliability, maintaining stable performance without degradation over $10^4$ endurance cycles. Furthermore, post-fabrication treatment was applied to achieve enhancement-mode operation. A depletion-load inverter was also demonstrated, exhibiting a voltage gain of 69.8 V/V. These results demonstrated the strong potential of LMP $InO_x$ nanosheet as a promising semiconducting channel in high-performance and power-efficient transistors for next-generation oxide electronics.

## METHODS

**Preparation of $InO_x$ nanosheet.** To exfoliate $InO_x$ nanosheet, indium metal (In, 99.995%) was used. A droplet of molten indium was placed on a substrate which was heated to 250 °C in ambient air. A second preheated substrate was then pressed on the molten indium under a uniaxial pressure of around 40 kPa, spreading the indium between two substrates. The pressure was held for 2 min to facilitate the spreading of indium native oxide, after which the two substrates were separated rapidly. After the pressing and separation procedures, the ultrathin $InO_x$ nanosheet was exfoliated onto two substrates. Following the separation procedure, a mechanical cleaning process was performed to remove residual indium. During this step, the substrates were kept at 250 °C while the surface was gently wiped using cotton buds dipped in ethanol, leaving a clean $InO_x$ nanosheet on the substrate. This cleaning step is laboratory-scale, and scalable standard solvent cleaning processes may be used for large-area printing.[29] No post annealing was performed in this process.

**Deposition of gate dielectrics.** $SiO_2/p^+$ Si wafers were purchased from WaferPro, where 300-nm $SiO_2$ was thermally grown via dry oxidation. A 30-nm $Al_2O_3$ layer was deposited on $p^+$ Si wafers at 250°C by PEALD (PicoSun Sunale) using Trimethylaluminium (TMA) as precursor and oxygen plasma as oxygen source. A 30-nm $HfO_2$ layer was deposited on $p^+$ Si wafers by PEALD (Ultratech Fiji G2) using Tetrakis(dimethylamido)hafnium(IV) (TDMAH) as precursor and oxygen plasma as oxygen source. The $HfO_2$ deposition was performed at the relatively low temperature of 150°C to maintain an amorphous phase, as polycrystalline films are known to result in increased roughness and higher leakage currents.[68, 69]

**Preparation of TEM samples.** TEM samples of cross-sectional layer stack of $Pt/InO_x/SiO_2$ were prepared using Focused Ion Beam (FIB) equipment (FEI Helios 600 NanoLab), where Pt layer was used for protection during ion milling. For plan-view TEM samples, typical mechanical thinning techniques were used, along with wet and dry etching processes. The $InO_x/SiO_2/p^+$ Si sample was firstly cut into ~2 mm × 2 mm square and mounted up on a supporting disk. The backside ($p^+$ Si side) was mechanically polished until the remaining thickness was less than 100 μm. A central dimple was then produced from the backside with a dimpling grinder, reducing the thickness in the target observation area to ~10-15 μm. The frontside ($InO_x$ side) was coated with Poly(methyl methacrylate) (PMMA) for protection, after which the sample was immersed in a solution of potassium hydroxide (KOH, 30 wt.%) under room temperature for ~3-4 hours to etch the Si and expose the $SiO_2$. Finally, ICP etching with $CHF_3$ plasma was performed to thin the exposed $SiO_2$ to less than 100 nm. The PMMA layer was subsequently removed by rinsing in acetone.

**Materials characterization.** The thickness and topography of $InO_x$ nanosheet were obtained by AFM (Bruker Dimension Icon) using a probe with nominal tip radius of 2 nm (Bruker SCANASYST-AIR). UV-Vis-NIR Spectrophotometer (Cary 5000) was used to capture UV−Vis spectroscopy of $InO_x$ on fused silica. GIXRD analysis was performed on $InO_x$ with a fused silica substrate using an X-ray Diffractometer (PANalytical X'Pert Pro). TEM (JEOL JEM-2100F) was used to investigate the cross-sectional layer stack of $Pt/InO_x/SiO_2$, as well as the plan-view observation. The thickness of thermally grown $SiO_2$, PEALD grown $Al_2O_3$, and $HfO_2$ were measured by using ellipsometer (JA Woollam M2000D).

**Device fabrication and electrical measurement.** Electron-beam lithography (RAITH 150) and photolithography (Heidelberg MLA 150) were used to define patterns for $InO_x$ semiconducting channel and metal deposition. RIE-ICP system (Samco 400 iP) was utilized to etch $InO_x$ with argon plasma. The ICP power and bias power were set to be 50 W and 100 W, respectively. E-beam/Thermal evaporator (Temescal BJD-2000) was adapted to deposit 40 nm of Ni and 20 nm of Au onto patterned samples to form metal contacts. The device parameters were confirmed by using SEM (FEI Verios 460).

**Post-fabrication treatment.** The RIE-ICP system (Samco 400 iP) was used to perform post-fabrication treatment on as-fabricated devices with oxygen plasma. The ICP power was set to be 100 W, while the bias voltage was kept at 0 W to minimize ion bombardment and ensure a mild, low-damage process. After 1 min of oxygen-plasma treatment, the devices were annealed at 120 °C for 20 min.

**Electrical measurement.** The electrical measurement of FETs and capacitors was finished by using a probe station (Signatone S-1160) connected to semiconductor device parameter analyzer (Keysight B1500A) and LCR meter (Keysight E4980A). All measurements were executed in ambient air.

## ASSOCIATED CONTENT

### Supporting Information

The following file is available free of charge.

Selected area electron diffraction (SAED) of $SiO_2$ and Pt protection layer, Energy-dispersive X-ray spectroscopy (EDX) of Pt/$InO_x$/$SiO_2$ stack, Capacitance-voltage (CV) measurements of gate dielectrics, Field-effect mobility of $InO_x$ FETs with different channel length, Scanning capacitance microscope (SCM) mapping, Electrical measurement of FETs with unpatterned $InO_x$, Statistical analysis of $InO_2$/$HfO_2$ FETs, Optical and SEM image of large-are $InO_x$ film, Electrical measurement of an $InO_x$/$SiO_2$ FETs with a partially interrupted channel, 3D AFM height profile of $InO_x$ before and after oxygen-plasma treatment, Cycle-to-cycle stability of the oxygen-plasma treated FET, Summary of reported indium oxide and doped-indium oxide transistors using different techniques, Summary of reported LMP transistors (PDF)

## AUTHOR INFORMATION

**Corresponding Authors**

**Shi-Rui Zhang** − Department of Electronic Materials Engineering, Research School of Physics, The Australian National University, Canberra, ACT 2601, Australia;

Email: shirui.zhang@anu.edu.au

**Robert Glen Elliman** − Department of Electronic Materials Engineering, Research School of Physics, The Australian National University, Canberra, ACT 2601, Australia;

Email: rob.elliman@anu.edu.au

**Authors**

**Sanjoy Kumar Nandi** − Department of Electronic Materials Engineering, Research School of Physics, The Australian National University, Canberra, ACT 2601, Australia

**Felipe Kremer** − Department of Electronic Materials Engineering, Research School of Physics, The Australian National University, Canberra, ACT 2601, Australia

**Shimul Kanti Nath** − School of Photovoltaic and Renewable Energy Engineering, University of New South Wales (UNSW Sydney), Kensington, NSW, 2052 Australia

**Wenzhong Ji** − Research School of Chemistry, The Australian National University, Canberra, ACT 2601, Australia

**Thomas Ratcliff** − Department of Electronic Materials Engineering, Research School of Physics, The Australian National University, Canberra, ACT 2601, Australia

**Li Li** − Australian National Fabrication Facility, Research School of Physics, The Australian National University, Canberra, ACT 2601, Australia

**Nicholas J. Ekins-Daukes** − School of Photovoltaic and Renewable Energy Engineering, University of New South Wales (UNSW Sydney), Kensington, NSW, 2052 Australia

**Teng Lu** − Research School of Chemistry, The Australian National University, Canberra, ACT 2601, Australia

**Yun Liu** − Research School of Chemistry, The Australian National University, Canberra, ACT 2601, Australia

## Author Contributions

S.-R.Z. was responsible for project design, implementation, materials characterization, preparation of TEM samples, device fabrication, data analysis and interpretation, and drafting the manuscript. S.K.N, T.R. R.G.E. provided project oversight and supervision, and assisted with data interpretation and drafting the manuscript. F.K. worked on TEM imaging, EDX mapping, SAED, and preparation of plane-view TEM samples. S.K.Nath and N.J.E.-D. undertook the deposition of $HfO_2$ layers through PEALD. W.J., T.L., and Y.L. worked on scanning capacitance microscopy and data analysis. L.L. prepared cross-sectional TEM samples via FIB. All the authors discussed the results and contributed to the final manuscript.

## Notes

The authors declare no competing financial interest.

## ACKNOWLEDGMENTS

The authors acknowledge access to NCRIS facilities and expertise at the ACT node of the Australian National Fabrication Facility (ANFF), the ANU ion-implantation Laboratory (iiLab), and the ANU Centre of Advanced Microscopy (CAM). S.-R.Z. thanks the China Scholarship Council and the Australian National University for scholarship support, and Dr. Azmira Jannat for her guidance in the early stage of this project. S.K.Nath thanks Dr. Elvin Mo for help with the ALD process. R.G.E. and T.R. acknowledge the Varian Semiconductor Division of Applied Materials and the Australian Research Council (ARC) (LP190100010) for financial support. T.L. and R.G.E. additionally acknowledge funding from ARC DP230100462, and T.L. and Y.L. acknowledge funding from ARC projects DE240100032, FL210100017, and DP250101852.

# Supporting Information

# High-Mobility Indium Native Oxide Transistors via Liquid-Metal Printing in Air

*Shi-Rui Zhang*[*,1], Sanjoy Kumar Nandi[1], Felipe Kremer[1], Shimul Kanti Nath[2], Wenzhong Ji[3], Thomas Ratcliff[1], Li Li[4], Nicholas J. Ekins-Daukes[2], Teng Lu[3], Yun Liu[3], and Robert Glen Elliman*[*,1]*

[1]Department of Electronic Materials Engineering, Research School of Physics, The Australian National University, Canberra, ACT 2601, Australia

[2]School of Photovoltaic and Renewable Energy Engineering, University of New South Wales (UNSW Sydney), Kensington, NSW, 2052 Australia

[3]Research School of Chemistry, The Australian National University, Canberra, ACT 2601, Australia

[4]Australian National Fabrication Facility, Research School of Physics, The Australian National University, Canberra, ACT 2601, Australia

[*]Corresponding authors: Shi-Rui Zhang; Robert Glen Elliman

Email: shirui.zhang@anu.edu.au; rob.elliman@anu.edu.au

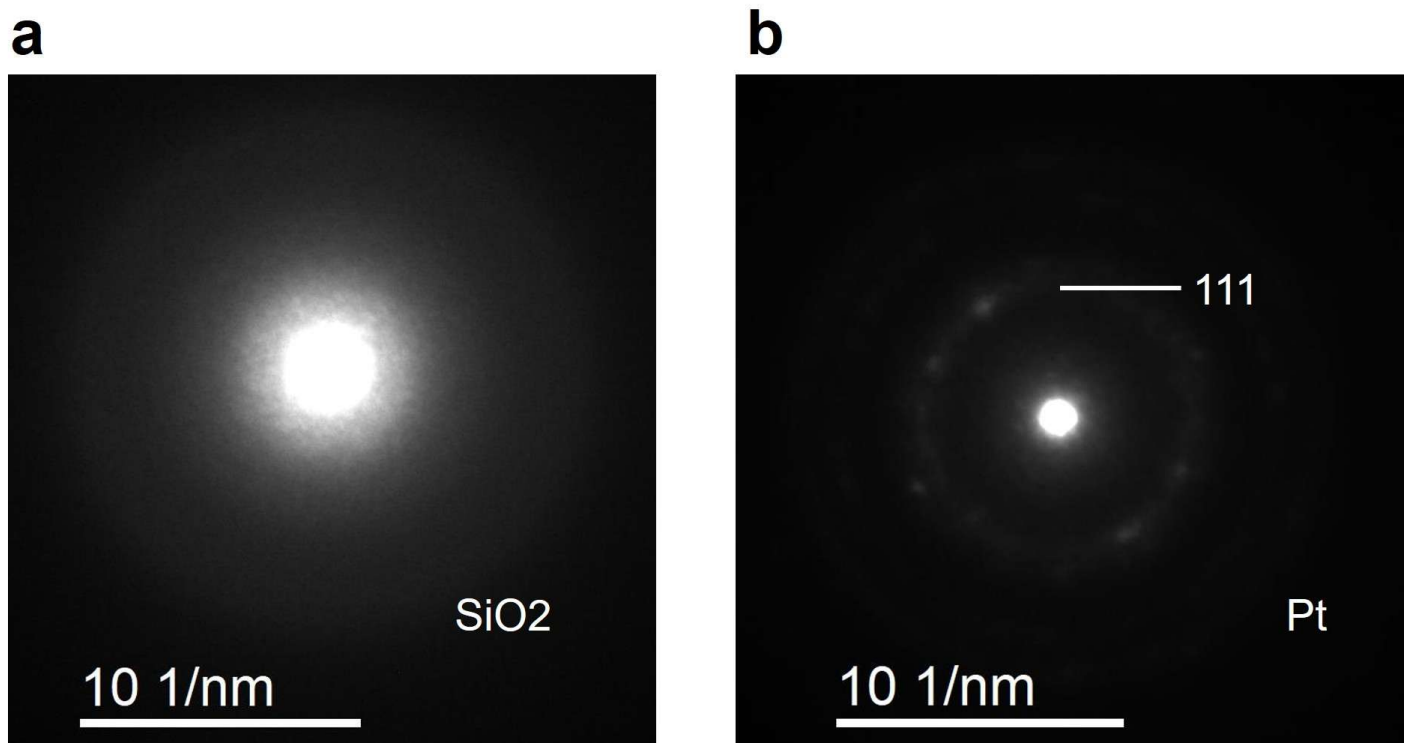

**Figure S1.** Electron diffraction patterns of $SiO_2$ and Pt protection layer. (a) Diffraction of $SiO_2$. (b) Diffraction of Pt protection layer. The typical diffraction ring is corresponding to (111) of nano-polycrystalline Pt.

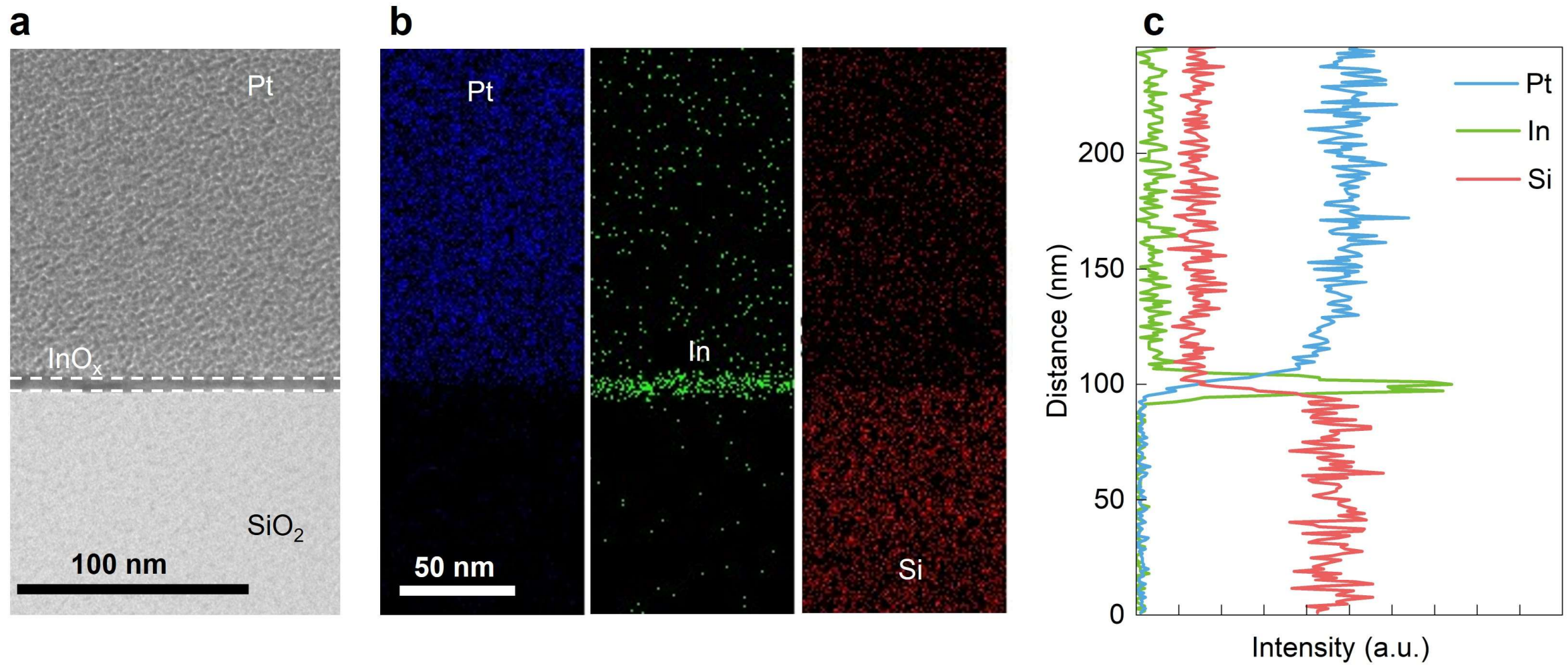

**Figure S2.** Energy-dispersive X-ray spectroscopy (EDX) of $Pt/InO_x/SiO_2$ stack. (a) Cross-sectional TEM image of $Pt/InO_x/SiO_2$ stack. (b) 2D elemental mapping of Pt, In, Si, respectively. (c) Corresponding EDX intensity distribution.

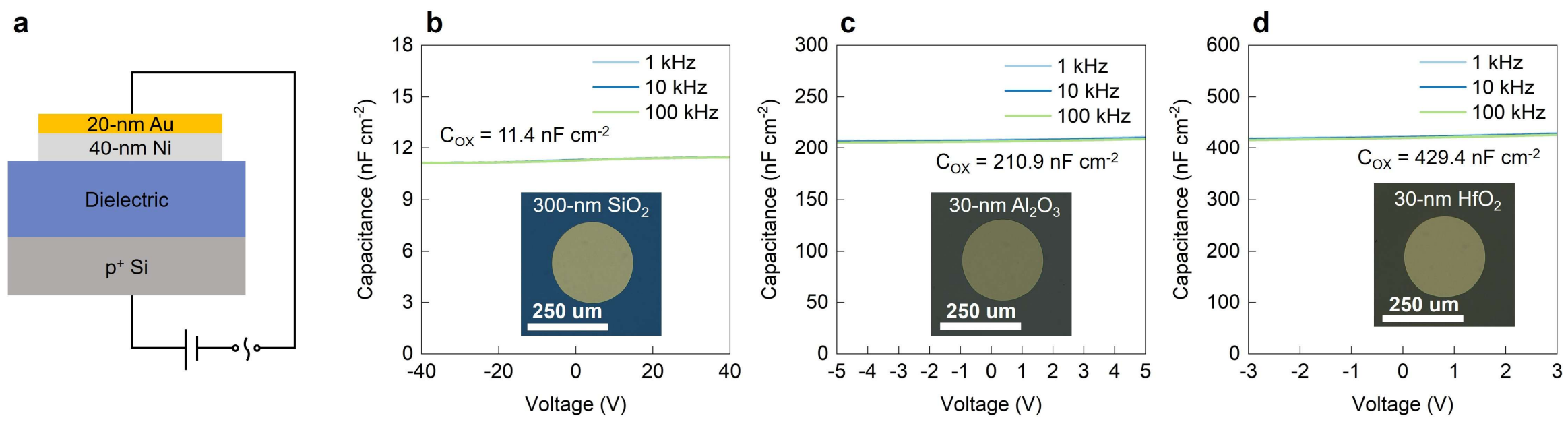

**Figure S3.** Capacitance-voltage (CV) measurements of gate dielectrics. (a) Schematic of capacitor structure and CV measurement. (b-d) CV curves of $SiO_2$, $Al_2O_3$, and $HfO_2$ capacitors, respectively. Insets: Optical images of metal/dielectric/$p^+$ Si capacitors.

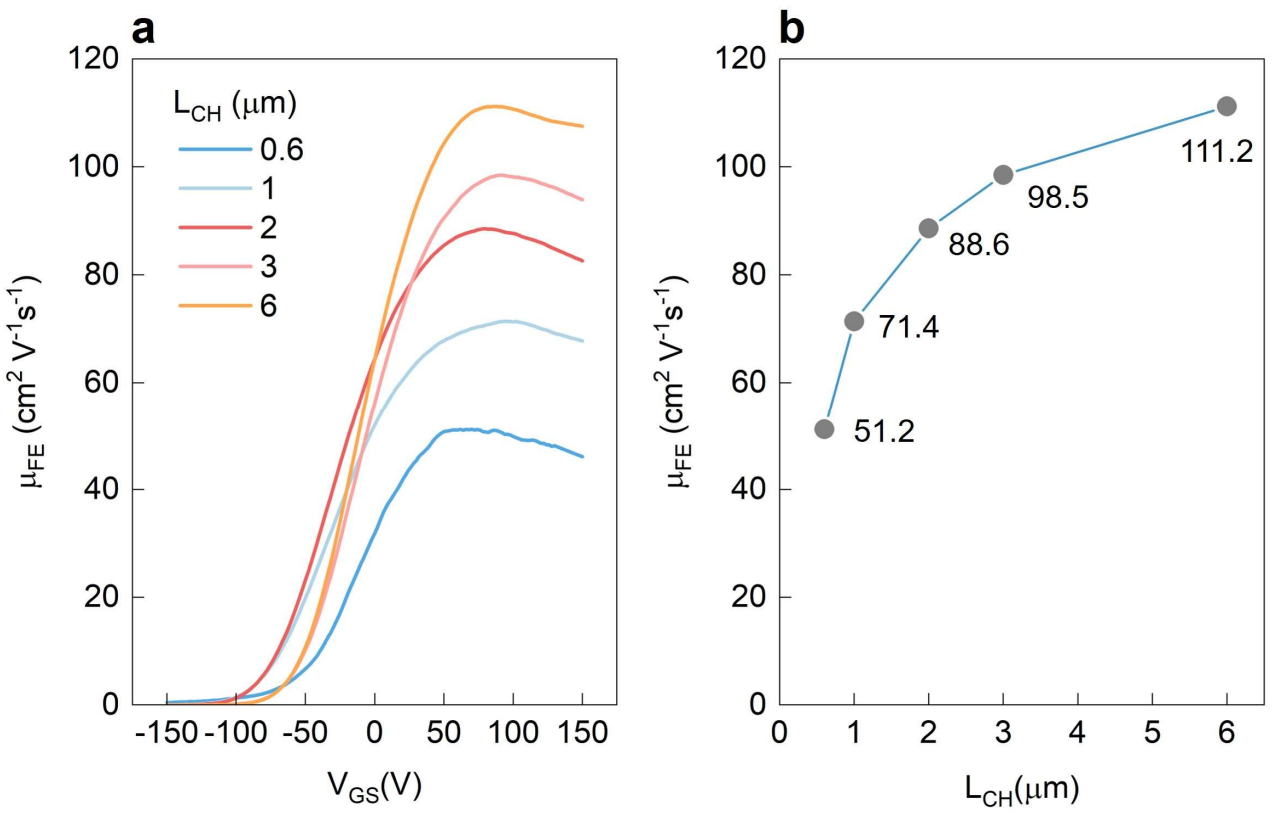

**Figure S4.** Field-effect mobility of $InO_x$ FETs with channel length of 0.6 μm, 1 μm, 2 μm, 3 μm, 6 μm, respectively. (a) $\mu_{FE}$ versus $V_{GS}$ plot. (b) Extracted maximum $\mu_{FE}$ of $InO_x$ FETs with different channel length.

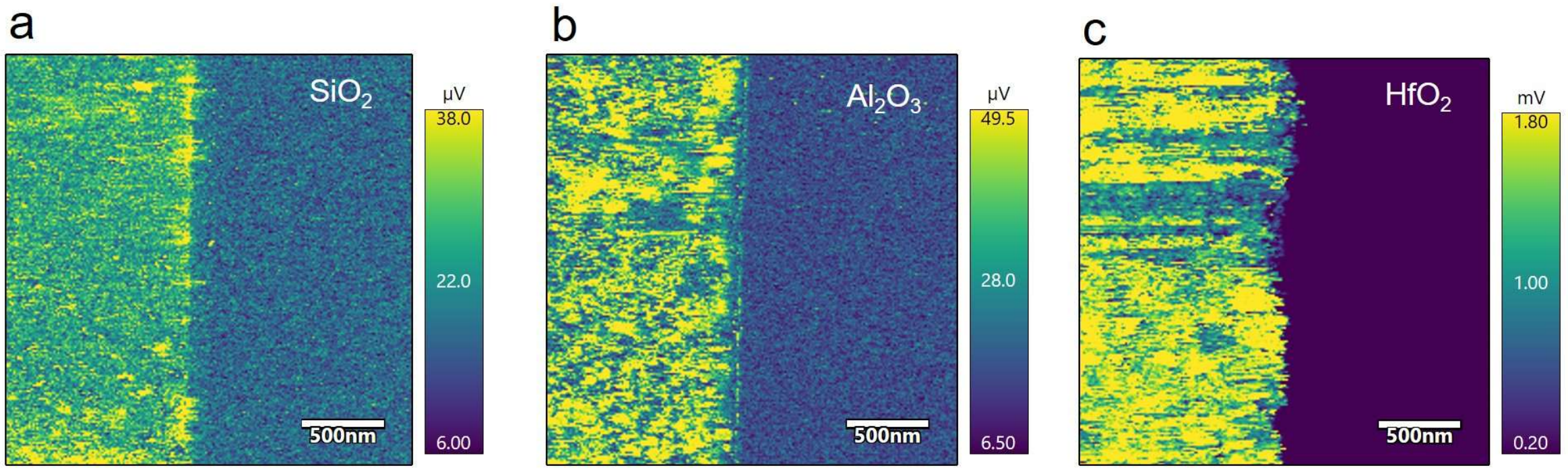

**Figure S5.** Scanning capacitance microscope (SCM) mapping. (a-c) SCM dC/dV amplitude mapping of samples with $InO_x$ on $SiO_2$, $Al_2O_3$ and $HfO_2$, respectively.

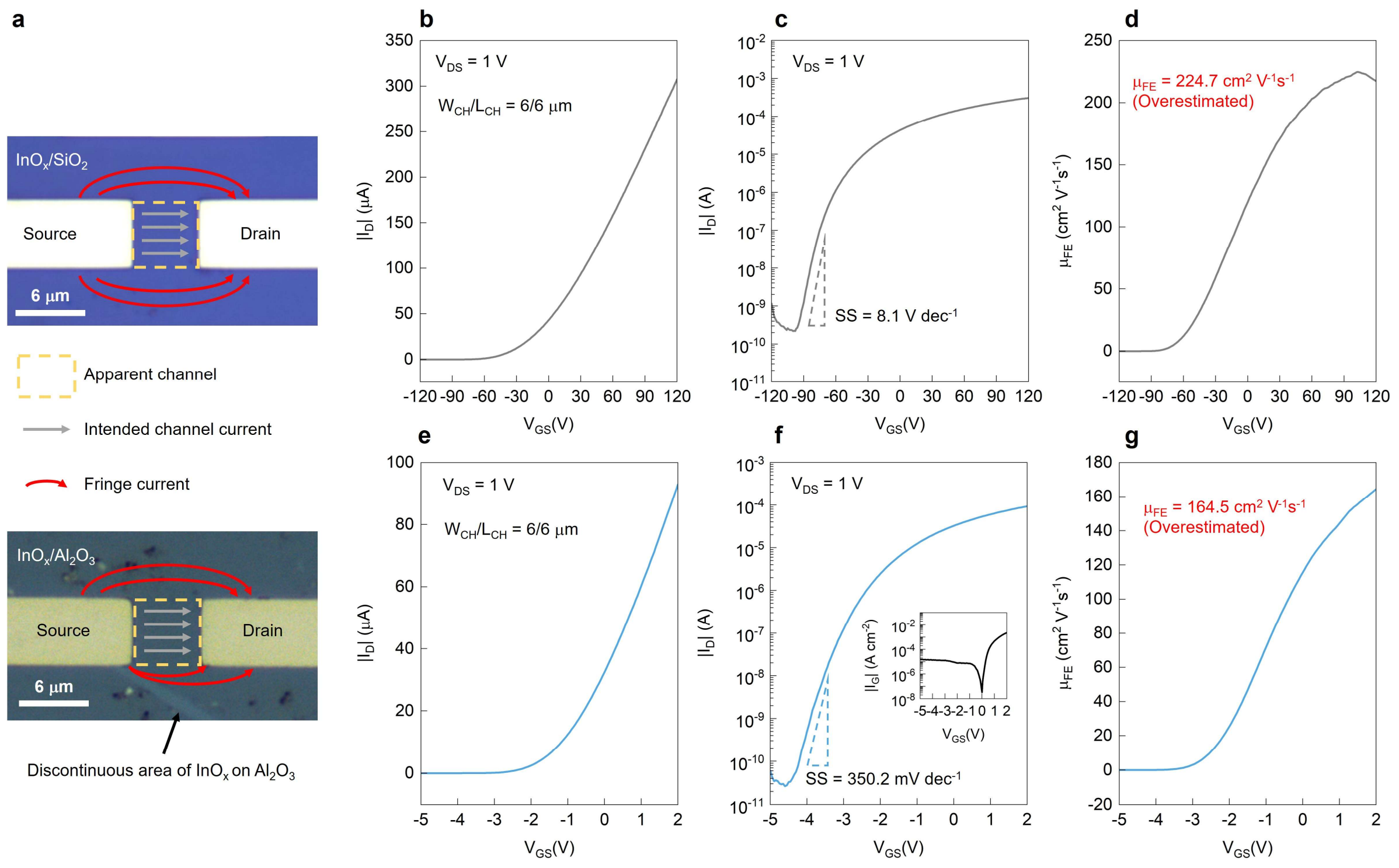

**Figure S6.** Electrical measurement of FETs with unpatterned $InO_x$. (a) Optical images of FETs and illustration of fringe effect. (b-d) Transfer ($I_D$-$V_{GS}$) curve in linear scale, Transfer curve in logarithmic scale and extracted field-effect mobility of FETs based on $SiO_2$ gate dielectric. (e-g) Transfer ($I_D$-$V_{GS}$) curve in linear scale, Transfer curve in logarithmic scale and extracted field-effect mobility of FETs based on $Al_2O_3$ gate dielectric. Inset in figure S6f: Gate leakage. An exponential increase of gate leakage current reflects electron transporting through trap sites inherently present in the gate dielectrics, as a result of a large contact area of semiconductor/dielectric interface

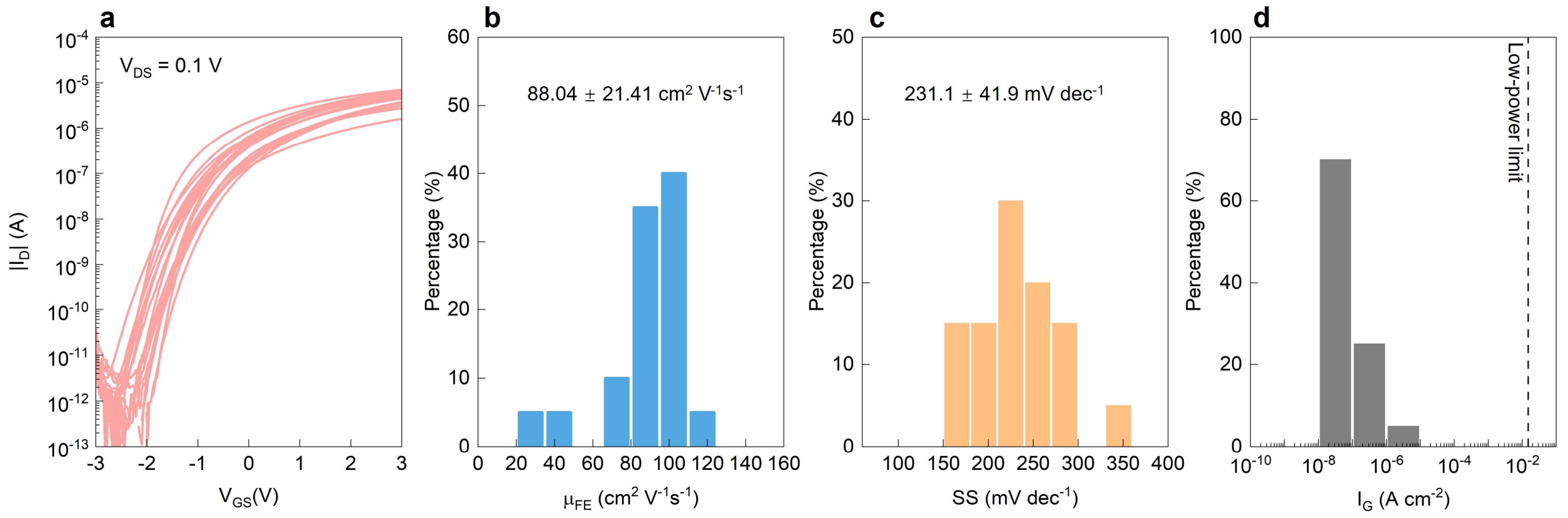

**Figure S7.** Statistical analysis of $InO_x/HfO_2$ FETs. (a) Transfer curves of 20 $InO_x/HfO_2$ FETs. Distribution of (b) $\mu_{FE}$ , (c) *SS*, (d) Gate leakage.

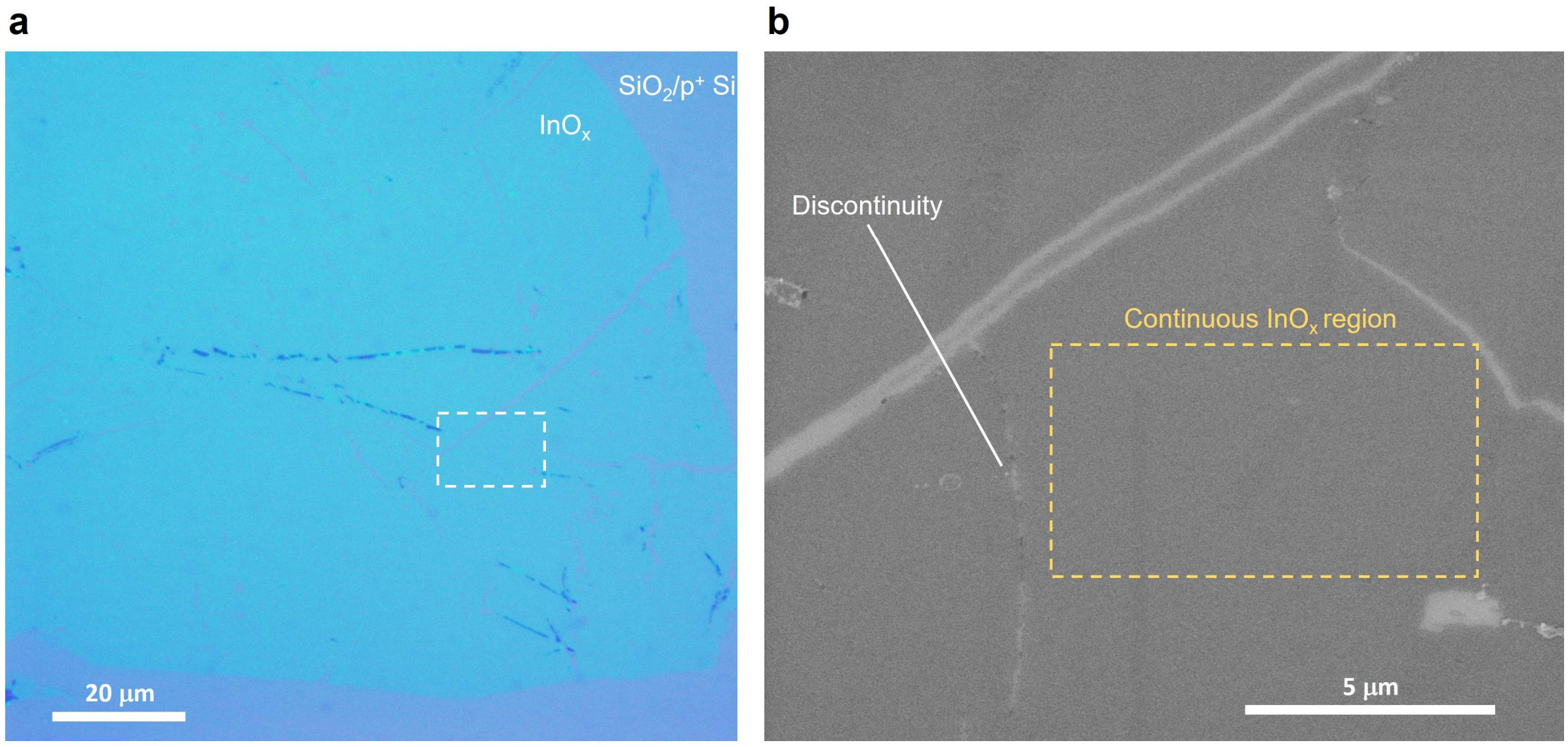

**Figure S8.** (a) Optical image of $InO_x$ film on $SiO_2/p^+$ Si substrate. (b) Magnified SEM image of $InO_x$ film within the white dashed square marked in (a). A typical $InO_x$ film spanning ~100μm exhibits visible discontinuities or local film breaks, some extending over several tens of micrometers. At higher magnification, additional sub-micrometer discontinuities that are not readily resolved optically become apparent in SEM when inspecting a magnified field of view.

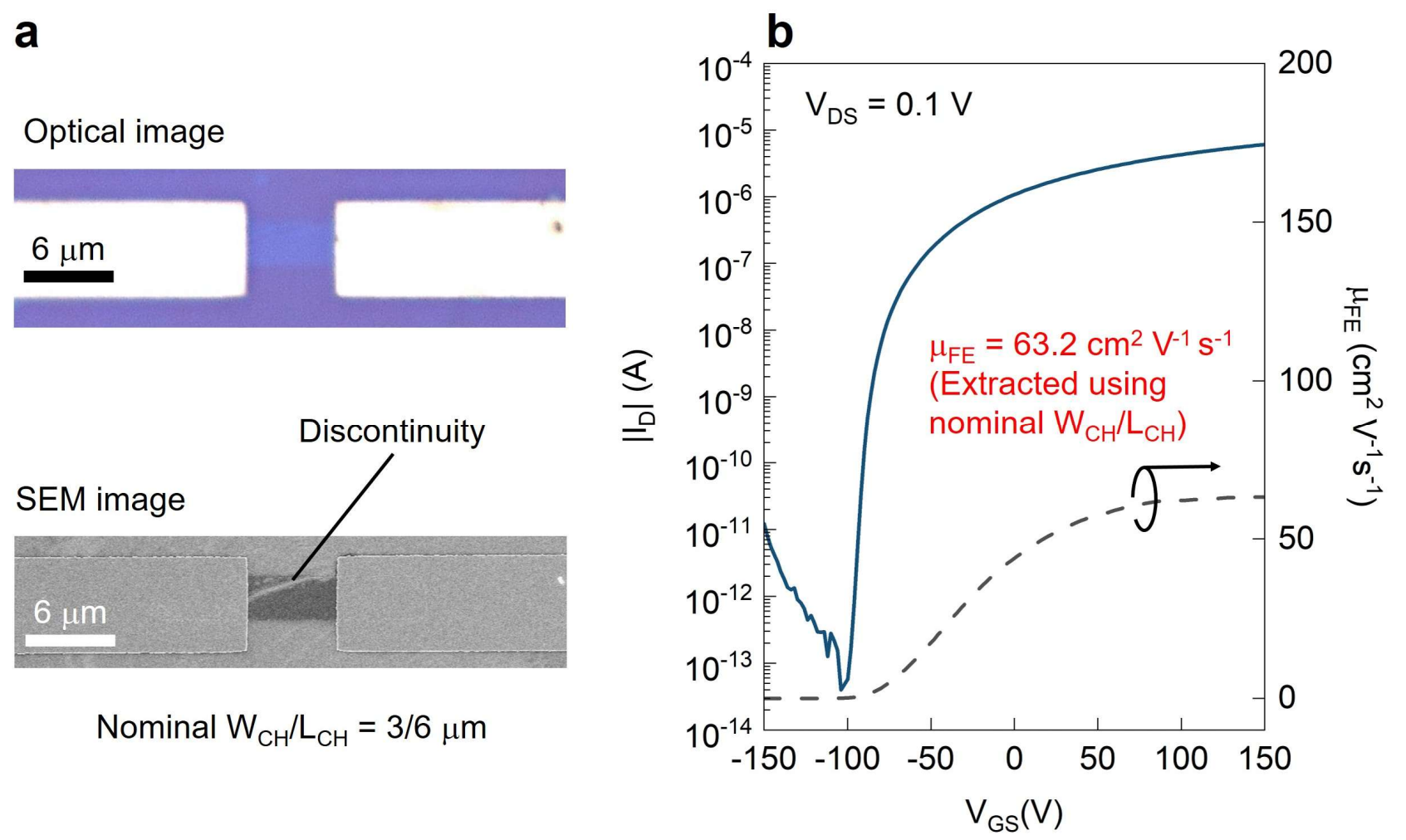

**Figure S9.** Electrical measurement of an $InO_x/SiO_2$ FETs with a partially interrupted channel. (a) Optical image and SEM image of an $InO_x/SiO_2$ FET with a discontinuity intersecting the defined channel region. (b) Transfer ($I_D$-$V_{GS}$) curves in logarithmic scale (left axis) and $\mu_{FE}$ versus $V_{GS}$ (right axis).

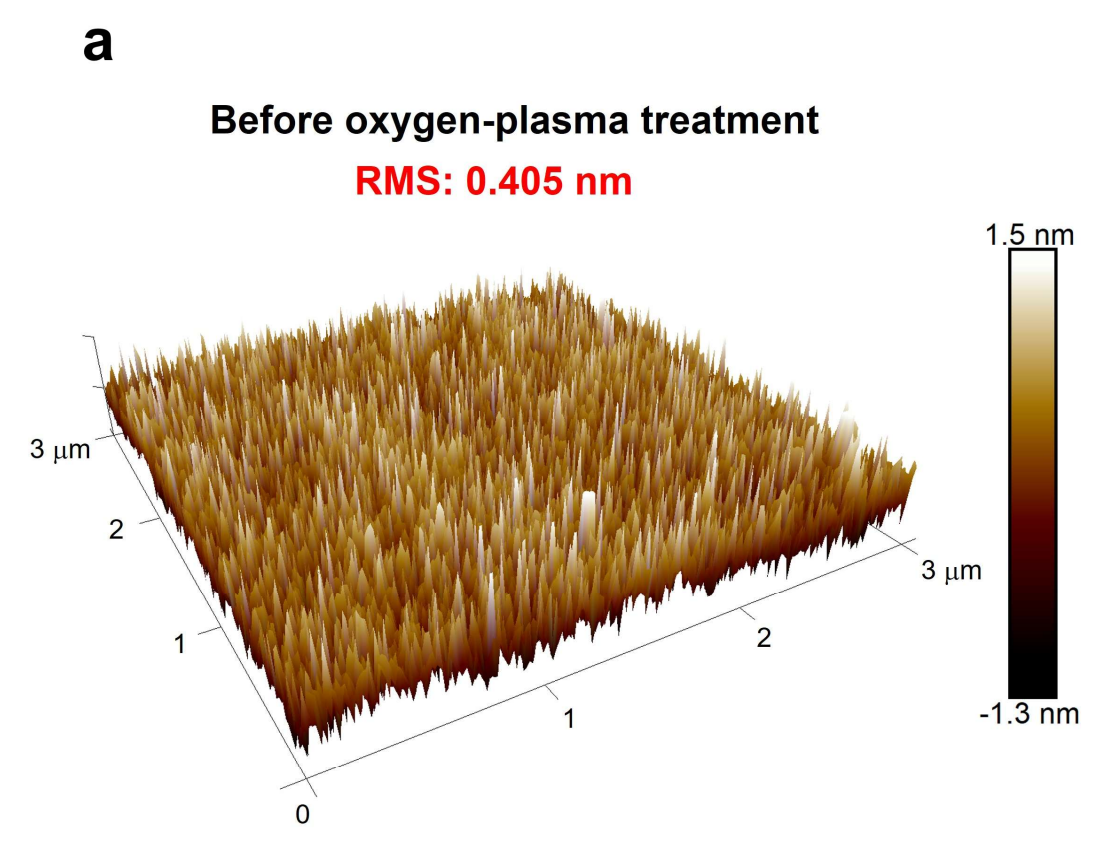

**Figure S10.** (a) 3D AFM height profile of $InO_x$ before oxygen-plasma treatment. (b) 3D AFM height profile of $InO_x$ layer after oxygen-plasma treatment. The measured RMS roughness is 0.405 nm before oxygen-plasma treatment and 0.426 nm after oxygen-plasma treatment, indicating a negligible change in surface roughness.

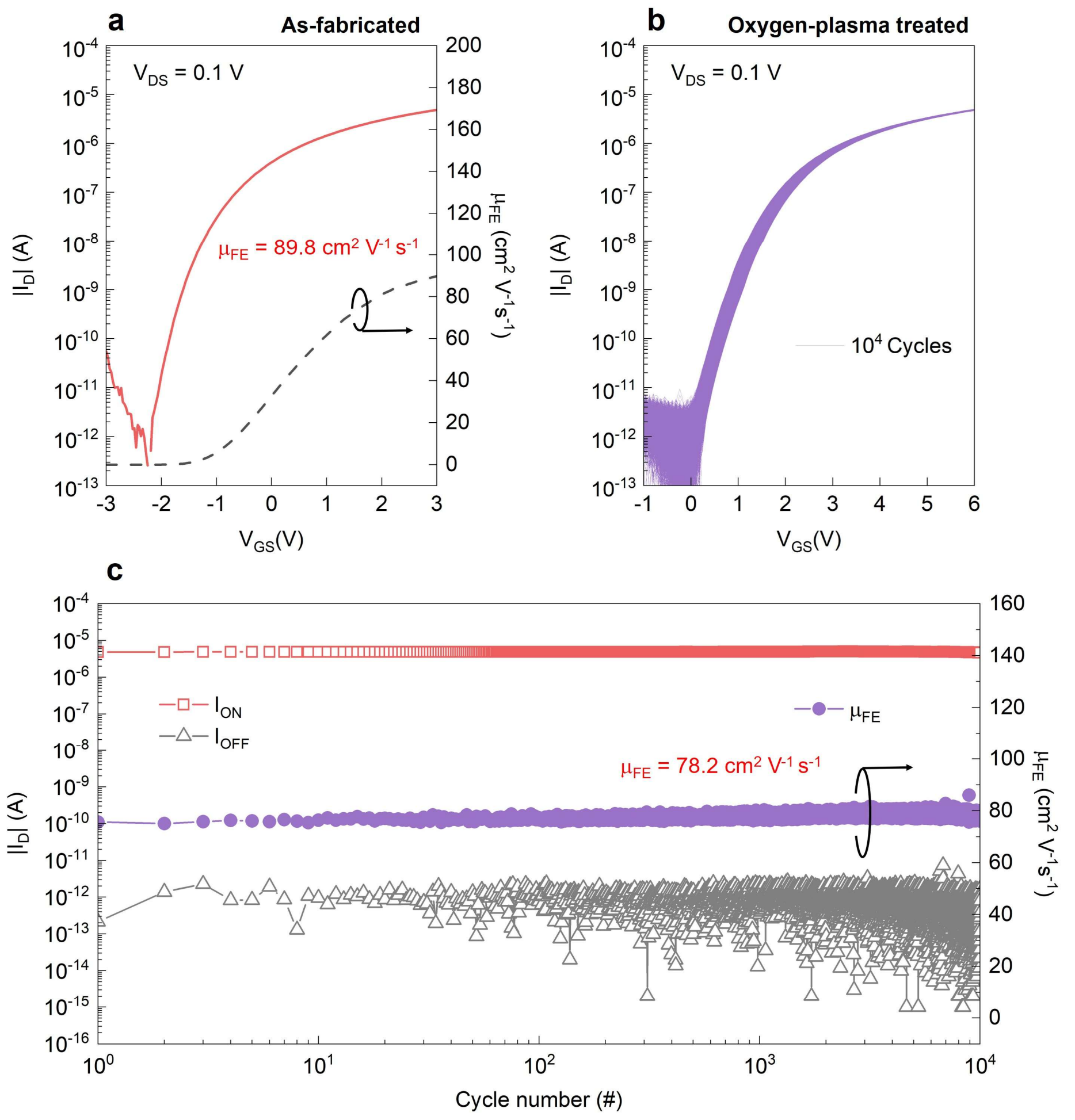

**Figure S11.** (a) Transfer curves of an as-fabricated FET and extracted $\mu_{FE}$ versus $V_{GS}$. (b) Consecutive transfer curves for $10^4$ cycles of the oxygen-plasma treated FET. (c) Cycle-to-cycle variation of $I_{ON}$, $I_{OFF}$ and $\mu_{FE}$. Before the oxygen-plasma treatment, the selected as-fabricated FET operates in depletion mode and exhibits an extracted $\mu_{FE}$ of 89.8 cm$^2$ V$^{-1}$ s$^{-1}$. After the plasma treatment, the FET shifts to enhancement-mode operation. Both $I_{ON}$ and $\mu_{FE}$ show cycle-to-cycle variability of < 2% over $10^4$ cycles, indicating that the plasma treatment does not measurably degrade the cycle-to-cycle stability. A slight decrease in $\mu_{FE}$ from 89.8 to 78.2 cm$^2$ V$^{-1}$ s$^{-1}$ is observed, which is consistent with previous studies on plasma-treated indium oxide transistors. [1] This reduction may be related to a reduced effective carrier concentration after plasma treatment. [2]

**Table S1.** Summary of reported indium oxide and doped-indium oxide transistors using different techniques

| Material | Method | Process[a)] Temperature (°C) | $L_{CH}$ (μm) | Gate dielectric | Mobility[b)] ($cm^2 V^{-1} s^{-1}$) | $V_{TH}$ (V) | $I_{ON}/I_{OFF}$ | *SS* ($V dec^{-1}$) | Year | Ref. |
|---|---|---|---|---|---|---|---|---|---|---|
| $In_2O_3$ | PLD | 350 | 350 | $SiO_2$ | 62.5 | -9.5 | $10^8$ | 0.36 | 2025 | [3] |
| $In_2O_3$: F | ALD | 300 | 10 | $SiO_2$ | 35.9 | 0.36 | —[c)] | 0.094 | 2024 | [4] |
| $In_2O_3$ | ALD | 400 | 20 | $HfO_2$ | 41.12 | -0.5 | — | 0.15 | 2024 | [5] |
| $In_2O_3$ | ALD | 225 | 0.2 | $HfO_2$ | 77 | — | — | — | 2021 | [6] |
| $In_2O_3$ | ALD | 350 | 350 | $SiO_2$ | 100 | -4.5 | $10^8$ | 0.38 | 2025 | [3] |
| $In_2O_3$ | ALD | 275 | 2.5 | $SiO_2$ | 104 | 0.5 | $>10^8$ | — | 2025 | [1] |
| $In_2O_3$ | ALD | 350 | 1 | $HfO_2$ | 113 | — | — | — | 2021 | [7] |
| $In_2O_3$ | ALD | 400 | 20 | $HfO_2$ | 116 | -0.12 | $>10^9$ | 0.066 | 2023 | [8] |
| $In_2O_3$ | ALD | 400 | 20 | $Al_2O_3$/ $HfO_2$ | 147.5 | 0.5 | — | 0.1 | 2025 | [9] |
| $In_2O_3$ | Sputter | RT | 150 | $SiO_2$ | 15.3 | 3.1 | $2.2\times10^8$ | 0.25 | 2010 | [10] |
| $In_2O_3$ | Sputter | RT | 120 | $SiO_2$ | 29.8 | 0.5 | $5.9\times10^6$ | 0.223 | 2019 | [11] |
| Tb: $In_2O_3$ | Sputter | 400 | 200 | $AlO_x$: Nd | 45 ($\mu_{SAT}$)[d)] | — | $10^8$ | — | 2022 | [12] |
| $InO_x$: H | Sputter | 250 | 1000 | $SiO_2$ | 125.7 | 0.16 | — | 1.22 | 2022 | [13] |
| $In_2O_3$: H | Sputter | 300 | 300 | $SiO_2$ | 139.2 | 0.2 | — | 0.19 | 2022 | [14] |
| $In_2O_3$ | Sol | 250 | 2000 | $Al_2O_3$ | 5.06 ($\mu_{SAT}$) | 4.43 | $3.6\times10^6$ | 0.3 | 2024 | [15] |
| $InO_x$ | Sol | 300 | 100 | $SiO_2$ | 13.95 | 0.67 | $1.42\times10^{10}$ | - | 2023 | [16] |
| LaInO | Sol | 500 | 250 | $SiO_2$ | 14.22 | 2.16 | $10^5$ | 0.84 | 2024 | [17] |
| $In_2O_3$ | Sol | 265 | 100 | $SiO_2$ | 35 | — | — | — | 2020 | [18] |
| $In_2O_3$ | Sol | 350 | 350 | $SiO_2$ | 40 | -5 | $10^8$ | 0.44 | 2025 | [3] |
| $In_2O_3$ | Sol | 400 | 100 | SAND | 43.7 | 2.2 | $10^6$ | 0.3 | 2008 | [19] |
| $In_2O_3$ | Sol | 300 | 100 | $AlO_x$ | 57.21 ($\mu_{SAT}$) | 0.57 | $6.0\times10^4$ | 0.22 | 2015 | [20] |
| $In_2O_3$ | Sol | 350 | 100 | $AlO_x$ | 127 | — | $10^6$ | 0.137 | 2013 | [21] |
| $InO_x$ | LMP | 250 | 6 | $Al_2O_3$ | 107.5 | 0.15 | $>10^7$ | 0.313 | This work | |
| $InO_x$ | LMP | 250 | 6 | $HfO_2$ | 107.2 | 0.31 | $>10^7$ | 0.204 | This work | |

[a)] Process temperature of the entire device fabrication process.

[b)] Unless otherwise specified, all values refer to field-effect mobility ($\mu_{FE}$).

[c)] Values not provided are marked as '—'.

[d)] $\mu_{SAT}$ is saturation mobility

**Table S2.** Summary of reported LMP transistors

| Channel Material | Type | $L_{CH}$ (μm) | Channel Patterning | Gate dielectric | EOT (nm) | Mobility[a)] ($cm^2 V^{-1} s^{-1}$) | $V_{TH}$ (V) | $I_{ON}/I_{OFF}$ | *SS* ($V dec^{-1}$) | Year | Ref. |
|---|---|---|---|---|---|---|---|---|---|---|---|
| SnO | p | 8 | —[b)] | $SiO_2$ | 80 | 0.7 | — | 20 | — | 2017 | 22 |
| SnO | p | 50 | $HNO_3$ wet etching | $SiO_2$ | 150 | 1 | -23 | $10^5$ | 4.9 | 2021 | 23 |
| β-$Ga_2O_3$ | p | 400 | — | $SiO_2$ | 500 | 21.3 | -3 | $10^4$ | — | 2019 | 24 |
| β-$TeO_2$ | p | 10 | — | $SiO_2$ | 300 | 232 | — | $>10^6$ | 0.103 | 2021 | 25 |
| $SnO_2$ | n | 10 | — | $SiO_2$ | 300 | 7.5 | — | $10^2$ | — | 2024 | 26 |
| β-$Ga_2O_3$ | n | 100 | $HNO_3$ wet etching | $SiO_2$ | 150 | 10.2 | 3.8 | $10^9$ | 0.163 | 2025 | 27 |
| α-$Bi_2O_3$ | n | 50 | — | $SiO_2$ | 150 | 2.2 ($\mu_{SAT}$)[c)] | 9.2 | $2.1 \times 10^6$ | 2.6 | 2024 | 28 |
| IGZO | n | 100 | — | $SiO_2$ | 500 | 14.25 | — | $>10^3$ | 0.46 | 2023 | 29 |
| IGO | n | 150 | HCl wet etching | $Al_2O_3$ | 23 | 6.5 | — | $10^5$ | 0.21 | 2024 | 30 |
| ITO | n | 10 | HCl wet etching | $Al_2O_3$ | 17.2 | 23.2 | -2.1 | $>10^9$ | 0.38 | 2022 | 31 |
| AIO | n | 200 | HCl wet etching | $SiO_2$ | 100 | 34.7 | -23 | $6 \times 10^3$ | — | 2025 | 32 |
| IAO | n | 20 | — | $SiO_2$ | 300 | 39.4 | — | $10^4$ | — | 2022 | 33 |
| IZO | n | 20 | — | $SiO_2$ | 300 | 87 | — | $10^5$ | — | 2021 | 34 |
| $InO_x$ | n | 50-500 | HCl wet etching | $SiO_2$ | 100 | 67.1 | 10 | $5 \times 10^6$ | — | 2022 | 35 |
| $In_2O_3$ | n | 20 | — | $SiO_2$ | 300 | 96 | — | $10^4$ | — | 2023 | 36 |
| $InO_x$ | n | 6 | ICP dry etching | $Al_2O_3$ | 16.4 | 107.5 | 0.15 | $>10^7$ | 0.313 | This work | |
| $InO_x$ | n | 6 | ICP dry etching | $HfO_2$ | 8 | 107.2 | 0.31 | $>10^7$ | 0.204 | This work | |

a) Unless otherwise specified, all values refer to field-effect mobility ($\mu_{FE}$).

b) Values not provided or processes not performed are marked as '—'.

c) $\mu_{SAT}$ is saturation mobility.